\documentclass[a4paper,11pt]{article}
\pdfoutput=1 

\usepackage{jinstpub} 

\usepackage{natbib}
\usepackage{lineno}
\usepackage{verbatim}
\usepackage{tabularx}
\newcolumntype{L}{>{\raggedright\arraybackslash}X}

\title{Design Development for the Beam Dump Facility Target Complex at CERN}

\author[a]{K. Kershaw,}
\author[a]{J.-L. Grenard}
\author[a,1]{M. Calviani,\note{Corresponding author.}}
\author[a]{C. Ahdida,}
\author[a]{M. Casolino,}
\author[b]{S. Delavalle,}
\author[b]{D. Hounsome,}
\author[a]{R. Jacobsson,}
\author[a]{M. Lamont,}
\author[a]{E. Lopez Sola,}
\author[b]{R. Scott,}
\author[a]{V. Vlachoudis,}
\author[a]{H. Vincke}

\affiliation[a]{European Organization for Nuclear Research (CERN)\\ Geneva, Switzerland}
\affiliation[b]{Oxford Technologies Ltd,\\Abingdon, England}

\emailAdd{marco.calviani@cern.ch}

\abstract{CERN has launched a study phase to evaluate the feasibility of a new high-intensity beam dump facility at the CERN Super Proton Synchrotron accelerator with the primary goal of exploring Hidden Sector models and searching for Light Dark Matter, but which also offers opportunities for other fixed target flavour physics programs such as rare tau lepton decays and tau neutrino studies. The new facility will require - among other infrastructure - a target complex in which a dense target/dump will be installed, capable of absorbing the entire energy of the beam extracted from the SPS accelerator. In theory, the target/dump could produce very weakly interacting particles, to be investigated by a suite of particle detectors to be located downstream of the target complex. As part of the study, a development design of the target complex has been produced, taking into account the handling and remote handling operations needed through the lifetime of the facility. Two different handling concepts have been studied and both resulting designs are presented.}

\keywords{Targets; Beam dumps; Radiation protection; Remote handling; Particle physics; Dark matter; Facility design}

\arxivnumber{1806.05920} 

\begin{document}
\maketitle
\flushbottom

\section{Introduction}
\label{sec:intro}

\subsection{The Beam Dump Facility}
\label{sec:intro:BDF}

CERN has launched a study phase to evaluate the feasibility of a new multi-purpose high-intensity facility with the primary goal of exploring Hidden Sector models and searching for Light Dark Matter, but which also offers opportunities for other fixed target flavor physics programs such as rare $\tau$ lepton decays and $\tau$ neutrino studies. The new facility will require - among other infrastructure - a target complex in which a dense target/dump (referred to in the rest of this document as the ``target") will be installed, capable of absorbing the entire energy of the beam extracted from the Super Proton Synchrotron (SPS) accelerator ($4\cdot 10^{13}$ protons/pulse, 1 s long slow extraction, $4\cdot 10^{19}$ Protons On Target (POT) per year, 355 kW average beam power).

The target (which will be dense to maximise production and to stop pions and kaons from propagating beyond the target) will be used to produce postulated very weakly interacting particles, to be investigated by a suite of particle detectors located downstream of the target complex. The objective is to search for portal interactions with hidden sector particles including Dark Matter candidates in a complementary way to colliders such as the LHC, i.e. looking for low mass particles with a very low coupling with normal matter. The first experiment that could make use of this facility will be the Search for Hidden Particle (SHiP) initiative~\citep{Anelli:2015pba,Alekhin:2015byh}.

\begin{figure}[htbp]
\centering %
\includegraphics[width=0.7\linewidth]{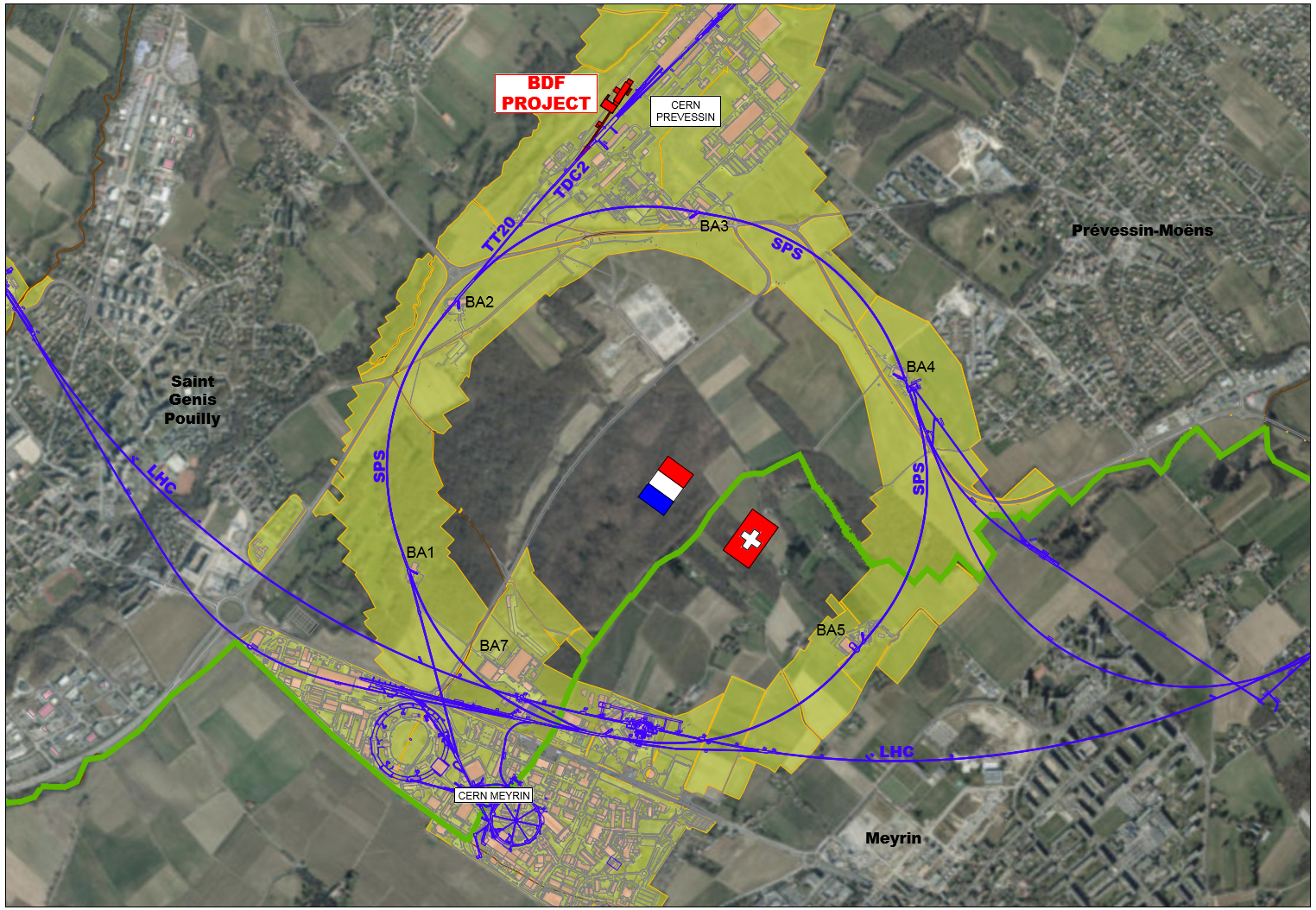}
\caption{\label{fig:BDFposition} Position of the BDF on CERN`s Prevessin site on the French site of the campus.}
\end{figure}

\begin{figure}[htbp]
\centering %
\includegraphics[width=0.7\linewidth]{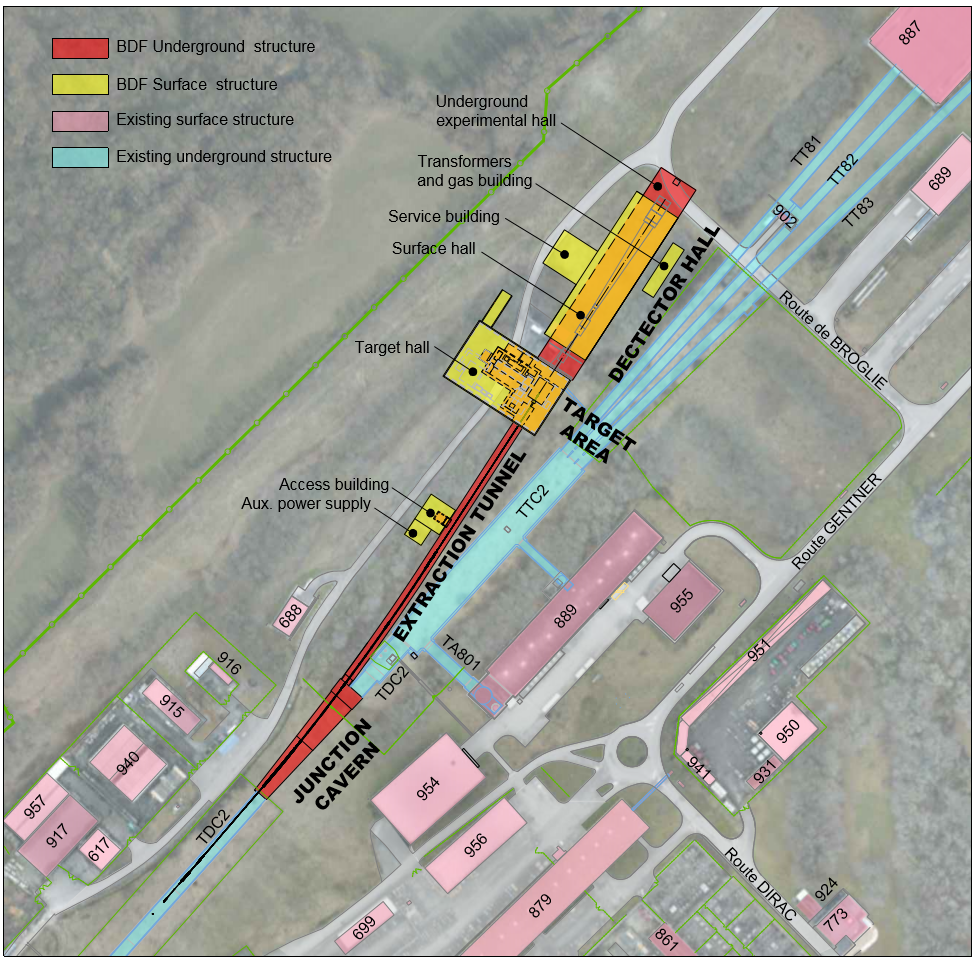}
\caption{\label{fig:BDFoverview} Plan view of the BDF extraction tunnel, target complex and detector hall next to the SPS ``North Area'' (North Area shown in blue).}
\end{figure}

A new junction cavern and extraction tunnel will be built to house the new beam line taking the proton beam from the SPS to the BDF target area. A large experimental hall is located immediately downstream of the target complex. It houses a muon shield \citep{Akmete:2017bpl} and the detector (Figure~\ref{fig:BDFposition} and Figure~\ref{fig:BDFoverview})~\citep{Anelli:2015pba,Alekhin:2015byh}.

\subsection{Introduction to the target complex}
\label{sec:intro:TTC}

The BDF production target (Section~\ref{sec:target}) will be at the heart of the new facility. High levels of radiation (both prompt and residual) will be produced by the SPS beam hitting the target (see Section~\ref{sec:RP}); a total cumulated dose near the target of around 500~MGy/year is expected. The target will be located in an underground area (to contain radiation as much as possible) located at about 15 metres below ground level. The depth of the infrastructure is determined by the location of the extraction line (TT20) which brings the beam from the SPS towards the target. The basic dimensions and layout of the target complex were determined as part of the conceptual design work for the SHiP design study~\citep{Calviani:SHIP}.

\begin{figure}[htbp]
\centering %
\includegraphics[width=1\linewidth]{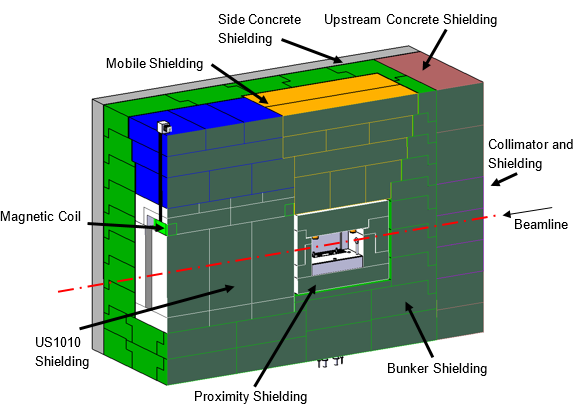}
\caption{\label{fig:BDFTTCcut} Cross-sectional view of the shielding around the target showing the magnetic coil and US1010 shielding yoke along with the different areas of the shielding inside the helium vessel (layout of shielding shown is for the ``crane'' concept described in Section~\ref{sec:crane}. Target not shown).}
\end{figure}

The target will be surrounded by approximately 3700 tonnes of cast iron and steel shielding with outer dimensions of around 6.8~m x 9.5~m x 8~m high (the so-called hadron absorber) to reduce the prompt dose rate during operation and the residual dose rate around the target during shutdown (Section~\ref{sec:TTCdesign}). The target and its surrounding shielding will be housed in a vessel containing gaseous helium slightly above atmospheric pressure in order to reduce air activation and reduce the radiation accelerated corrosion of the target and surrounding equipment (Figure~\ref{fig:BDFTTCcut}). References to similar target complex infrastructures for conventional neutrino beams can be found at \citep{Abe:2011ks,Papadimitriou:2017ytl}.

The SPS beam will enter the helium vessel through a beam window, then pass through a collimator which serves to protect the target and adjacent equipment from misalignment of the incident SPS beam and to protect the equipment in the extraction tunnel from particles (essentially neutrons generated by the target) travelling backwards relative to the incident beam. Downstream of the target a magnetic coil and US1010 steel yoke are used to produce a magnetic field of 1.5-1.6 T in order to sweep high energy muons produced in the target to reduce experimental backgrounds~\citep{Anelli:2015pba}. The secondary beam will pass through the downstream helium vessel wall and leave the target complex via a 5 cm thick steel plate ``window'' at the upstream end of the detector hall.

The target and the shielding immediately around it will be water cooled. All the shielding in the helium vessel will be built up of blocks; the layout and geometries of which are designed to avoid direct radiation shine paths and to minimise the number of block movements needed to allow exchange of failed equipment. A helium purification system is foreseen to allow flushing of air to ensure a 99.9\% pure helium atmosphere in the vessel after initial installation and after maintenance interventions. Ventilation of the target complex will use a cascade of pressure between different zones to provide containment of any radioactive contamination that could potentially be released according to ISO standard 17873~\citep{ISO17873}.

Remote handling and manipulation of the target and surrounding shielding will be mandatory due to the high residual dose rates. The target complex has been designed to house the target and its shielding in the helium vessel along with the cooling, ventilation and helium purification services below ground level. The target complex design allows for removal and temporary storage of the target and shielding blocks in the cool-down area below ground level and includes dedicated shielded pits for storage of the highest dose rate equipment.

A 40-tonne capacity overhead travelling crane in the target complex building will be used for initial installation and will carry out the handling and remote handling of the shielding blocks and other equipment as needed for assembly and maintenance of the facility. All the shielding blocks and other heavy equipment are designed to be compatible with the crane's lifting capacity. Target and beam elements are to be aligned within +/-10 mm with respect to the incoming proton beam.

\section{BDF target design}
\label{sec:target}

The BDF target is at the core of the installation and can be considered as a beam dump, since it is designed to safely absorb the full 400 GeV/c SPS primary beam. Its conception has been optimised also from a physics perspective (in terms of geometry, material, gaps, etc.) to maximise the production of charmed mesons. The high power deposited on the target is one of the most challenging aspects of the BDF target design, with 320 kW average power and 2.3 MW over the 1-second spill.

\begin{figure}[htbp]
\centering %
\includegraphics[width=0.8\linewidth]{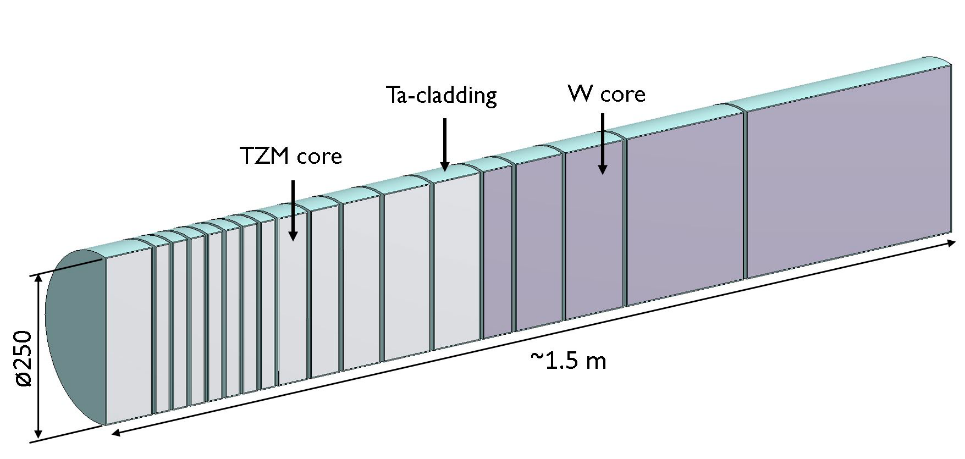}
\caption{\label{fig:BDFtarget} Layout of the BDF target absorbing blocks.}
\end{figure}

The target material selection is based on the requirement of having high-Z materials with a short nuclear interaction length, with the aim of increasing the re-absorption of pions and kaons produced (contrary to a neutrino-producing target). The proposed target design consists of several collinear cylinders of TZM ((0.08\%)~titanium-(0.05\%)~zirconium-molybdenum alloy) and pure tungsten (W), cladded with pure Ta or a W-containing Ta-alloy. The target cylinders are 250 mm in diameter and have different lengths, from 25 to 80 mm for the TZM blocks and from 50 to 350 mm for the pure tungsten blocks. The length of the target cylinders has been iteratively adjusted to minimise the level of temperatures and stresses reached in the target, for a total target length of around 1.5 m (Figure~\ref{fig:BDFtarget}).

Given the high energy deposited and the high temperatures reached during operation, the target requires active water cooling. The cooling water will flow through 5 mm gaps between the different blocks with high velocity (around 5 m/s on average), in order to provide an effective heat transfer between the cooling medium and the target blocks. 

In order to avoid undesired corrosion/erosion effects induced by the high-speed water in contact with pure W and TZM, all the blocks are cladded via diffusion bonding achieved by means of Hot Isostatic Pressing (HIP) with either Ta or Ta2.5W-alloy, due to their high erosion/corrosion resistance and convenience as high-Z material~\citep{LANSCEcladding,ISIScladding}.

\begin{figure}[htbp]
\centering %
\includegraphics[width=1\linewidth]{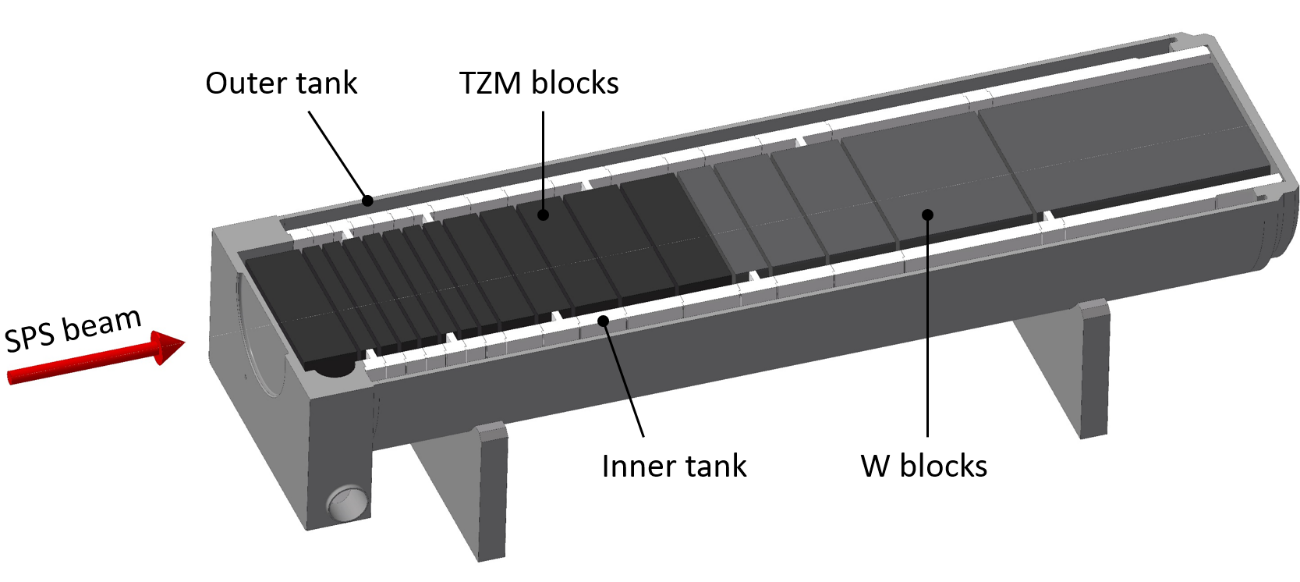}
\caption{\label{fig:BDFtargetassembly} BDF target assembly preliminary design, showing the absorbing blocks and the inner vessel.}
\end{figure}

The preliminary design of the target assembly includes two concentric tanks: the outer tank will ensure the leak-tightness of the assembly and provide an interface for the water and electrical connections, while the inner tank will support the target blocks and enclose the target cooling circuit (Figure~\ref{fig:BDFtargetassembly}).

A BDF target prototype has been built to be tested at CERN under a slow extracted proton beam. The target replica consists of a reduced scale prototype with the same block thickness distribution but a reduced diameter and will allow cross-checking of the thermo-mechanical calculations performed and will provide an insight of the material response under beam irradiation. Beam testing of the target is foreseen during late 2018. The design of the BDF target and of the target prototype will be the subject of a dedicated paper.

\section{Radiation protection aspects of the BDF target complex}
\label{sec:RP}

As the BDF aims to push the primary proton beam to a power of up to 355 kW, radiation protection considerations strongly determine the design of the facility. In compliance with the applicable CERN radiation protection rules regarding doses to personnel and environmental impact, the relevant radiological aspects have been carefully addressed at the design stage of all components.

\begin{figure}[htbp]
\centering %
\includegraphics[width=0.6\linewidth]{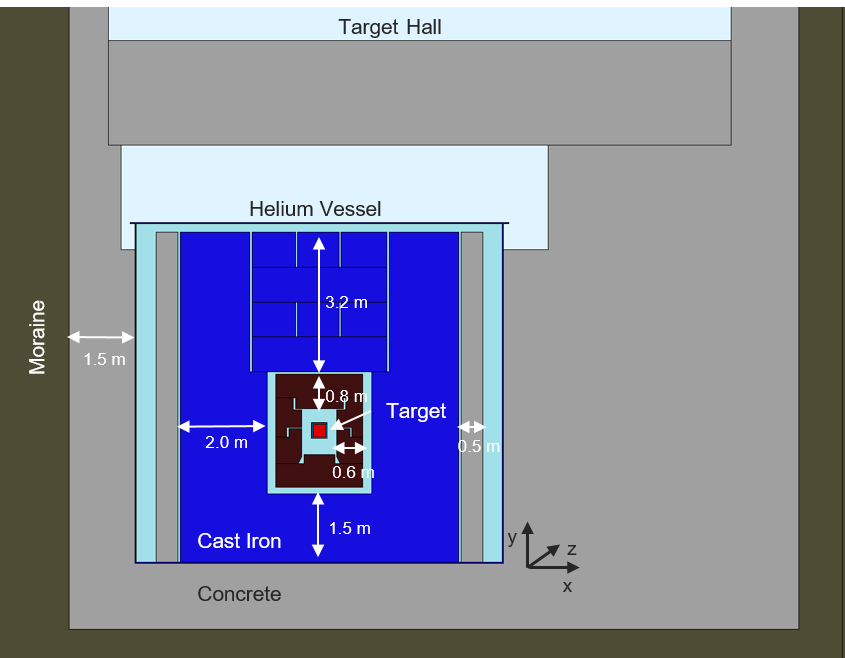}
\caption{\label{fig:FLUKARPmodel} Cross-sectional cut through the BDF target complex FLUKA geometry. The target is displayed in red, the proximity shielding in dark red, the bunker shielding in blue, concrete in grey, and helium in light-blue.}
\end{figure}

A particular radiation consideration is the high prompt and residual dose rates that call for considerable shielding (roughly 600~m$^3$) and remote interventions in the target area. The shielding was consequently designed with the objective to keep the various radiological hazards originating from the operation of the BDF facility as low as reasonably possible, while taking the constraints from the different stages of the experiment, i.e.~the construction, operation, maintenance and dismantling, into account. The envisaged configuration is such as to avoid activation of the fixed concrete civil engineering structures simplifying not only the dismantling/decommissioning but also possible changes of scope of the installation. Figure~\ref{fig:FLUKARPmodel} shows a cross-sectional view of the BDF target complex geometry, modeled with the FLUKA Monte Carlo code~\citep{BOHLEN2014211}.

\begin{figure}[htbp]
\centering %
\includegraphics[width=0.8\linewidth]{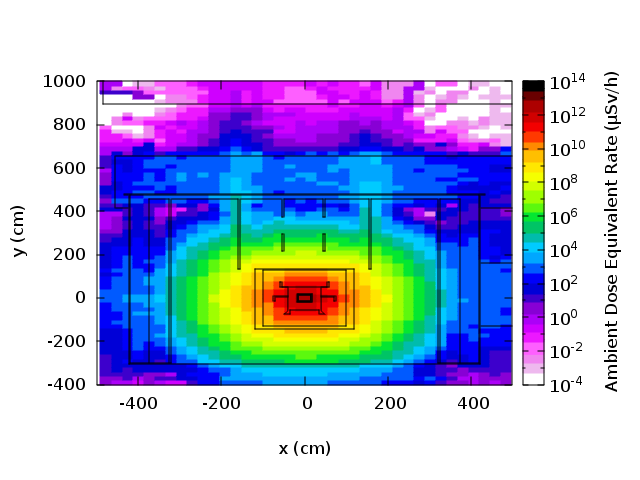}
\caption{\label{fig:FLUKARPresults1} Cross-sectional cut through the BDF target complex showing prompt radiation levels in $\mu$Sv/h. Values close to the target reach 10$^{6}$~Sv/h, while on top of the helium-vessels - thanks to the massive cast iron shielding - values are in the order hundreds of mSv/h.}
\end{figure}

Figure~\ref{fig:FLUKARPresults1} depicts the expected prompt dose rates in the accessible area around the BDF target, as evaluated with the FLUKA code. On top of the helium vessel prompt dose rates reach hundreds of mSv/h, while in the target hall, where controlled access will be allowed during beam operation, the prompt dose rates are designed to be <1~$\mu$Sv/h. It was furthermore verified that the prompt dose rates originating from the facility do not affect the surrounding experimental areas and that the stray radiation at any point outside the fenced CERN site does not cause annual effective doses to any member of the public exceeding the strict limits in place.

\begin{figure}[htbp]
\centering %
\includegraphics[width=0.8\linewidth]{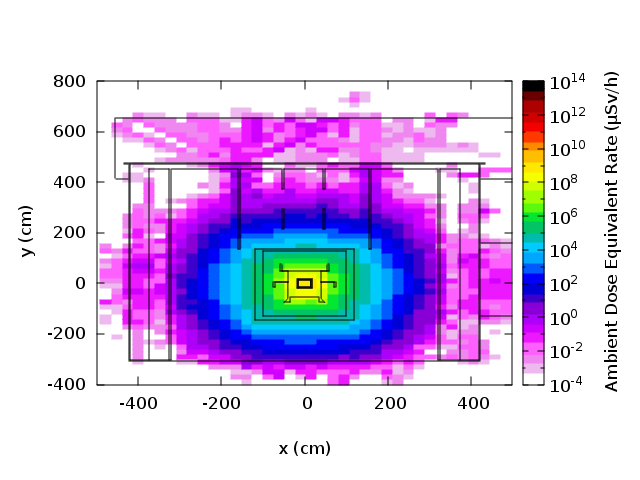}
\caption{\label{fig:FLUKARPresults2} Residual dose rates in the BDF target complex for a cooling time of 1 week in $\mu$Sv/h. After 5 years of irradiation and 1 week of cooling time, the residual dose rates are in the order of hundreds of Sv/h around the target.}
\end{figure}

The expected residual dose rates around the BDF target are shown in Figure~\ref{fig:FLUKARPresults2}. After 5 years of irradiation and 1 week of cooling time, the residual dose rates are in the order of hundreds of Sv/h around the target and the proximity shielding requiring full remote handling and designated storage areas for handled elements. The iron blocks are specially designed and optimised for remote handling, as they will become highly activated. The closest area relative to the target where human intervention is required is above and next to the helium vessel enclosing the shielding. Here, maximum residual dose rates of a few $\mu$Sv/h after 1 week of cooling down are reached with the helium vessel closed and all shielding elements in place.

In addition, the risk and environmental impact from releases of radioactivity by air and water as well as the soil activation heavily influences the design. To minimise air activation the target and the iron shielding are placed in a helium atmosphere with a maximum of 0.1\% air contamination. The use of helium is motivated by the fact that it only gives rise to the formation of tritium, which has a significantly lower radiological impact than the radionuclide arising from air. To keep soil and ground water activation to negligible levels, the iron and concrete shielding surrounding the target was optimised. In addition to design considerations based on normal operation, failure scenarios and critical interventions were taken into account in the design phase.

\section{Target complex design methodology}
\label{sec:TTCdesign}

\subsection{Design study stages}
\label{sec:TTCdesign:design_study}

After the initial work to determine the main requirements and basic layout of the target complex as explained in the introduction, the target complex design was further developed by going into more detail on the handling and remote handling operations required throughout the life of the facility. This work aimed to demonstrate the feasibility of the construction, operation, maintenance of the BDF target complex along with decommissioning of the key elements. 
The remote handling of highly activated radioactive objects, such as target, beam window (Section~\ref{sec:common_elements:beam_window}), collimator (Section~\ref{sec:common_elements:collimator}), shielding blocks and magnetic coil (Section~\ref{sec:common_elements:magnetic_shielding}), along with their connection and disconnection within the target complex building was studied by CERN in collaboration with Oxford Technologies Ltd in order to arrive at the integrated designs presented here. In addition to ``foreseen'' remote handling operations, such as target exchange, the study considered ``unforeseen'' remote handling operations needed to recover from failures or damage to equipment (Section~\ref{sec:recovery}). The study included the conceptual design of lifting, handling and remote handling equipment for the highly activated objects along with the necessary water, helium and electrical connections compatible with the radiation environment and remote handling constraints. These designs were then integrated in to the target complex as a whole.

\subsection{The two handling concepts studied}
\label{sec:TTCdesign:handling_concepts}

Target complex designs based on two different handling concepts have been developed: the ``crane concept'' (Section~\ref{sec:crane}) and the ``trolley concept'' (Section~\ref{sec:trolley}). The crane concept relies on the overhead travelling crane (Section~\ref{sec:common_elements:trav_crane}) in the target complex building for the movements of the target, shielding and magnetic coil during the life of the facility. The trolley concept has the target and its main services installed on a mobile trolley running on rails allowing quicker access to this critical component. 
For both concepts the core elements (target internals, beam window (Section~\ref{sec:common_elements:beam_window}), helium vessel (Section~\ref{sec:common_elements:helium}), collimator (Section~\ref{sec:common_elements:collimator}), magnetic coil and US1010 shielding (Section~\ref{sec:common_elements:magnetic_shielding})) are essentially the same and are common to the target complex designs produced for both concepts. The main differences between the crane and trolley concepts are in the way the target and water-cooled proximity shielding are supported, installed and removed from the helium vessel and how their services are connected and disconnected.

\section{Elements common to both handling concepts}
\label{sec:common_elements}

\subsection{Beam window}
\label{sec:common_elements:beam_window}
\begin{figure}[htbp]
\centering 
\includegraphics[width=.32\textwidth]{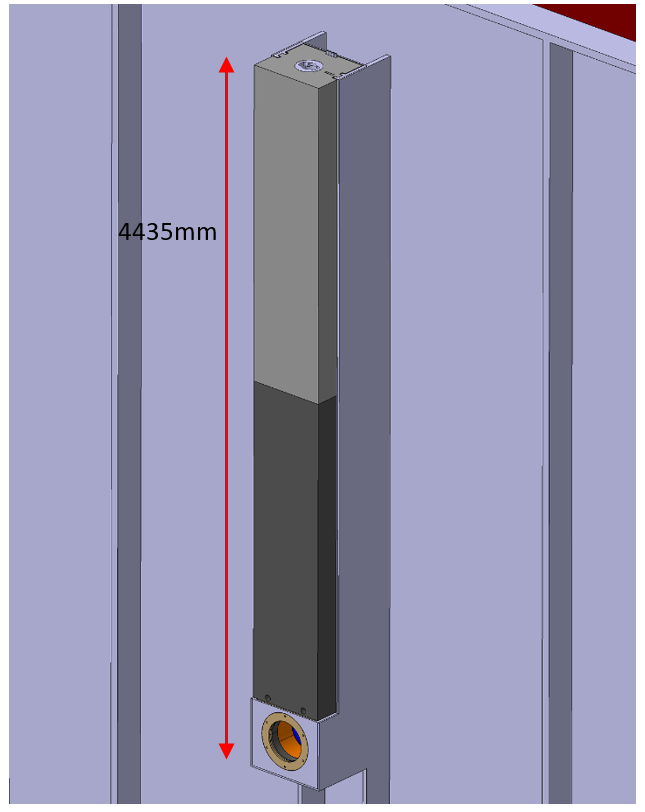}
\qquad
\includegraphics[width=.6\textwidth,origin=c]{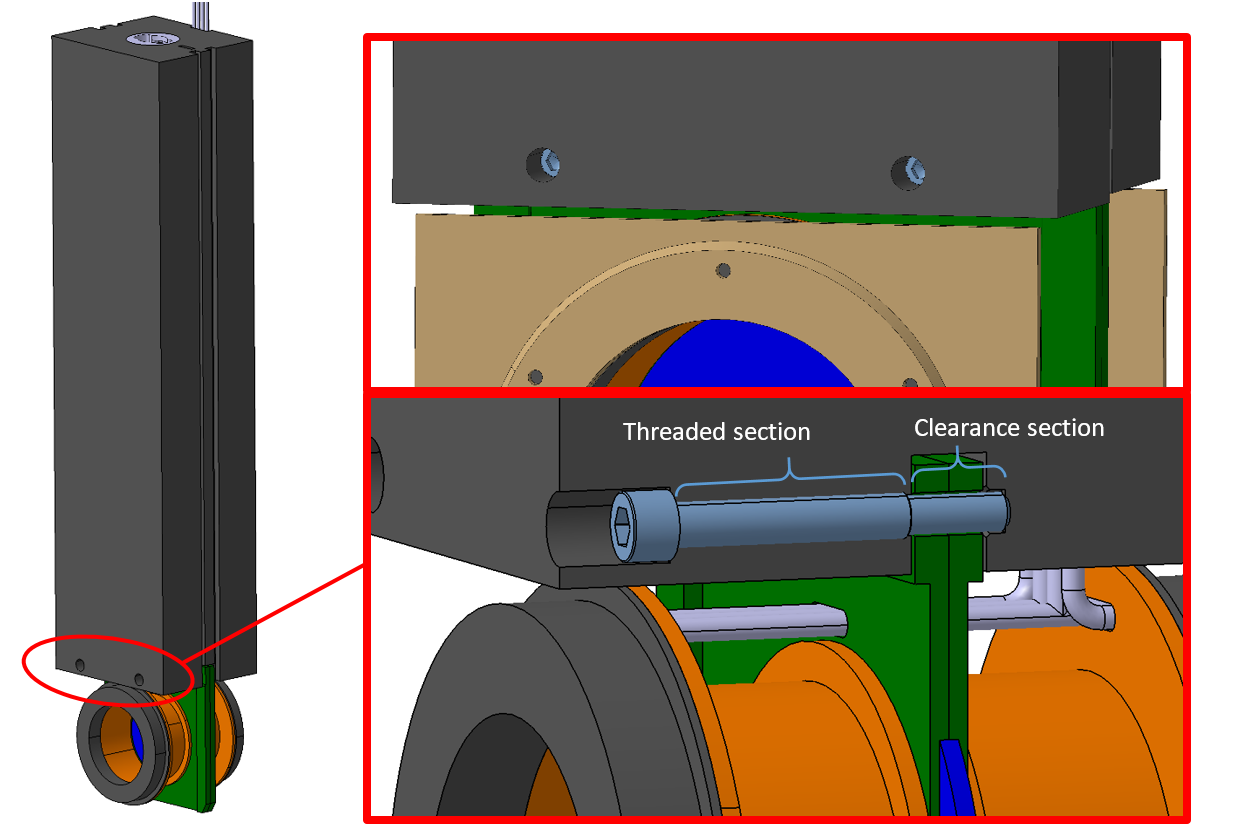}
\caption{\label{fig:beam_window} The proton beam window. {\it (Left)} Beam window assembly installed on the helium vessel. {\it (Center)} Lower shielding block with the beam window attached. {\it (Right)} Details of the remotely operable attachment of the beam window to the lower shielding block.}
\end{figure}

The beam window is mounted on the outside of the helium vessel (Section~\ref{sec:common_elements:helium}). It uses two inflatable pillow seals to seal it to the upstream beam pipe on one side, and to the helium vessel on the other.  The beam window assembly is designed to allow remote replacement using the crane. First the crane lifts out an upper shielding block and then a lower shielding block with the window attached to it. The blocks are transferred to the remote handling section of the cool-down area where the beam window is then disconnected from the lower block using a pair of through-the-wall master-slave manipulators (Figure~\ref{fig:beam_window}).

\subsection{Collimator}
\label{sec:common_elements:collimator}
\begin{figure}[htbp]
\centering %
\includegraphics[width=0.65\linewidth]{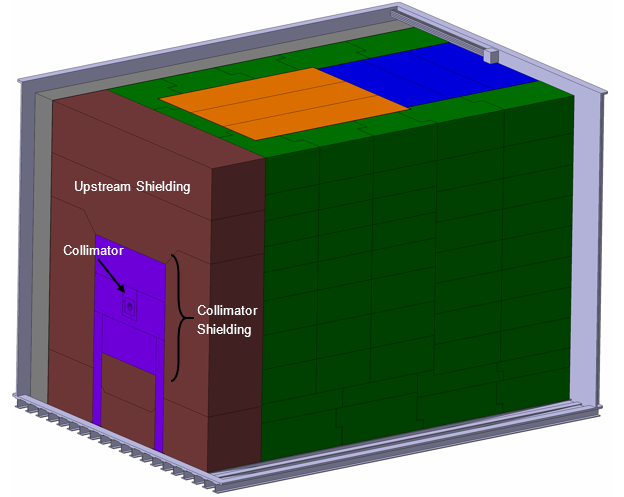}
\caption{\label{fig:collimator} The collimator and surrounding shielding in cut-away helium vessel.}
\end{figure}

The collimator is formed as part of a removable block within the upstream shielding within the helium vessel (Figure~\ref{fig:collimator}). It consists of a 150~cm long, 20~cm diameter graphite mask employed to protect the downstream target and shielding assembly against beam misalignment. It can be remotely installed and removed by the crane of the target hall. The shielding immediately around the collimator is supported on pillars to ensure accurate alignment.

\subsection{Magnetic coil and US1010 shielding}
\label{sec:common_elements:magnetic_shielding}
\begin{figure}[htbp]
\centering 
\includegraphics[width=.4\textwidth]{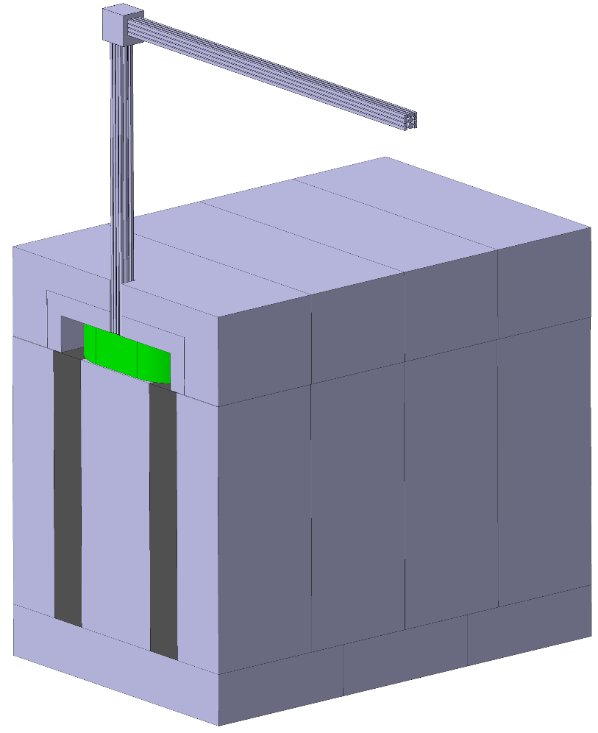}
\qquad
\includegraphics[width=.4\textwidth,origin=c]{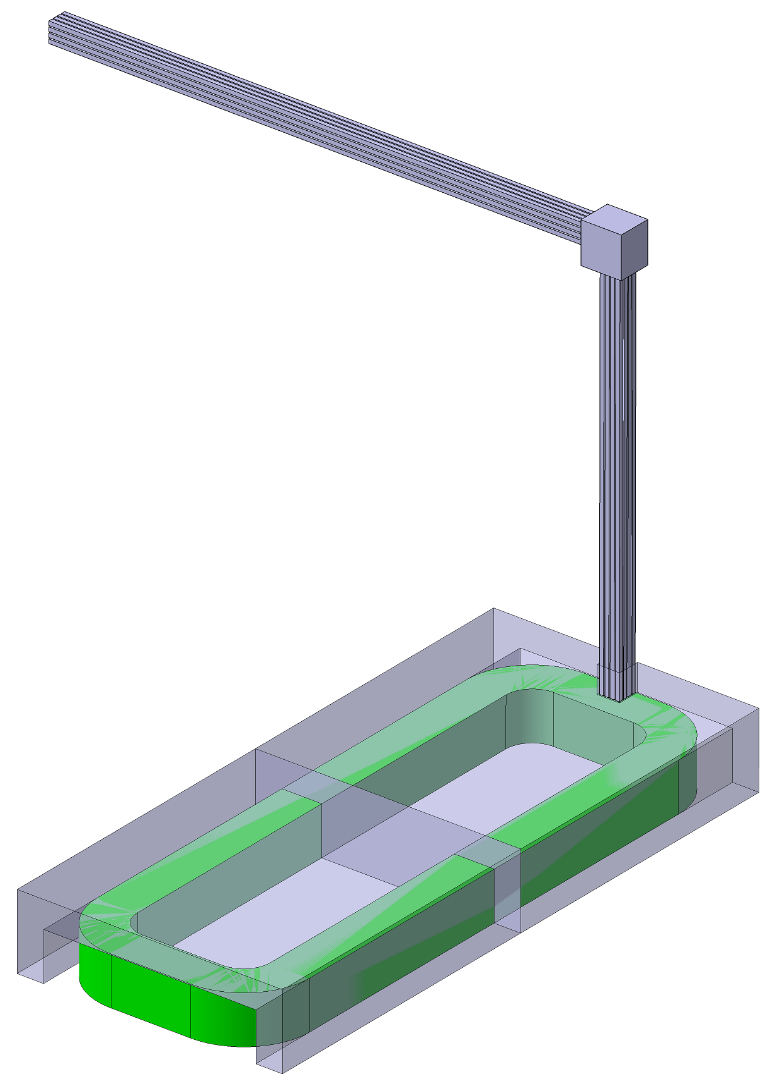}
\caption{\label{fig:magnetic_coil} Magnetic coil and US1010 shielding. {\it (Left)} US1010 shielding that is downstream of the target and the magnetic coil. {\it (Right)} Magnetic coil with its surrounding shielding which restrains the coil during operation and provides lifting point for the crane. The junction box between the horizontal and vertical portion of the services connections represents the connections which are made and broken by hands-on interventions.}
\end{figure}

The magnetic coil and US1010 shielding are designed to work together to provide a magnetic field downstream of the target of 1.5-1.6 T in order to sweep the muons produced in the target away from the detector acceptance to reduce experimental backgrounds. The shielding includes two sections of non-magnetic stainless-steel blocks (shown in black in Figure~\ref{fig:magnetic_coil}) to ensure that the magnetic field is correctly guided through the steel yoke.

The coil is lifted in and out of the helium vessel along with its surrounding US1010 shielding and its services connections which protrude above the top layer of the shielding in the helium vessel. The location of this service connection allows for hands-on connection and disconnection above the bunker shielding (when all of the blocks are installed).  The design and layout of the US1010 blocks have been optimised to give the minimum number of gaps seen by the magnetic field generated by the coil whilst remaining compatible with achievable manufacturing and handling precision.

\subsection{Helium vessel and services}
\label{sec:common_elements:helium}
\begin{figure}[htbp]
\centering %
\includegraphics[width=1\linewidth]{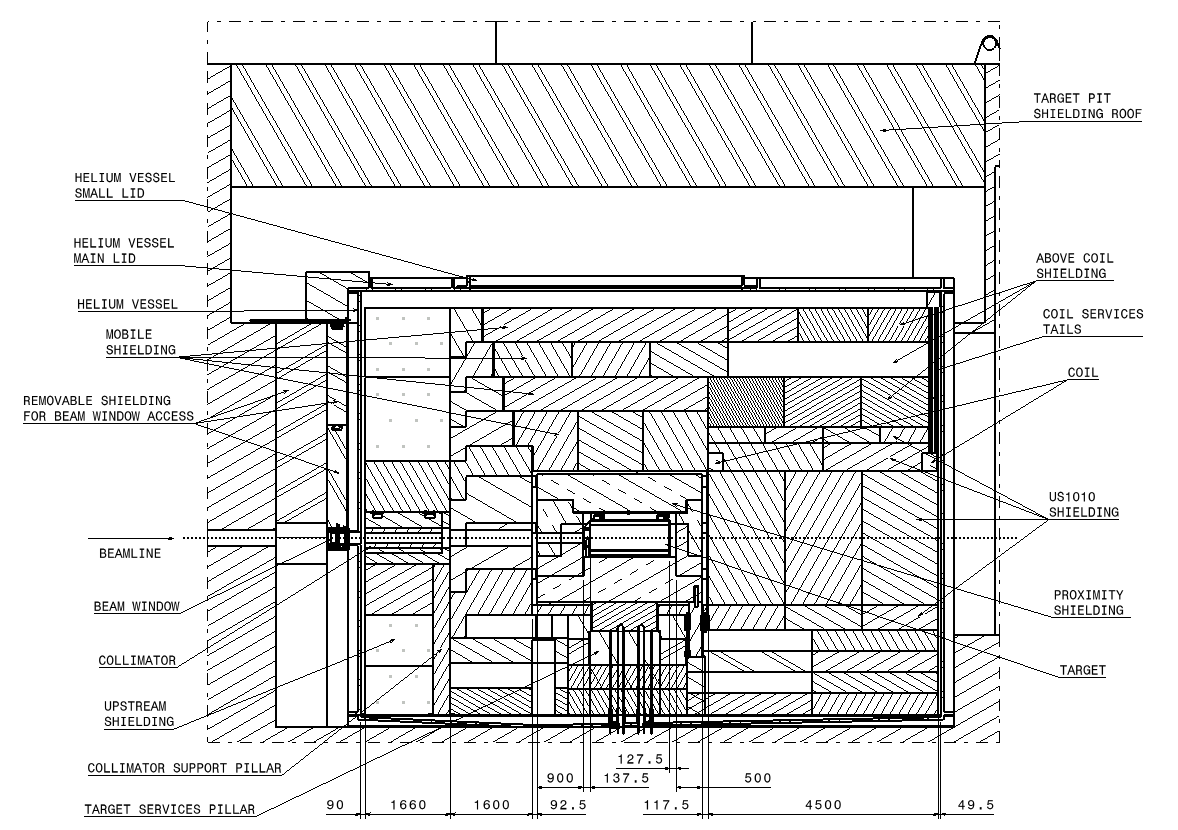}
\caption{\label{fig:He_vessel_services} Section view of helium vessel in the target pit showing the main elements and different shielding areas (crane concept shown).}
\end{figure}

The helium vessel (Figure~\ref{fig:He_vessel_services}) has to support the loads due to the shielding inside it, ensure good helium leak tightness, allow the passages of services, permit flushing of air with helium and also draining of any water leaks. 

To allow for flushing of helium and draining of any water leaks, the floor of the vessel slopes down to a drain point. This drain point is connected to a sump room so that any water leaking into the vessel drains even in the event of pump failure. Service galleries are included in the civil engineering structure to allow connections between the cooling and ventilation (Section~\ref{sec:common_elements:CV}) and sump rooms. The structural design of the helium vessel will be carried out as part of a design phase which is outside the scope of the current paper.  

The helium vessel designs for the two handling concepts have some differences in the position of service passages and access openings for target and shielding movements. In the crane concept design, the services (water, helium, electrical) for the target and proximity shielding enter the helium vessel from below, whilst the services for the magnetic coil enter from the vessel side close to the top (Figure~\ref{fig:He_vessel_services}).

\subsection{Cool-down and remote handling areas}
\label{sec:common_elements:cool_down}
A cool-down area below ground level is provided for temporary storage of the target, shielding, magnetic coil, beam window etc. The area has been designed to allow temporary storage and transfer as required, for all foreseen maintenance operations during the life of the facility. 

In addition, a remote handling area, equipped with a pair of through-the-wall master-slave manipulators, has been included as part of the cool-down area to carry out operations such as the disconnection of the beam window from its shielding block, removal of the magnetic coil from its surrounding shielding and unforeseen repair work if needed.

\subsection{The building overhead travelling crane}
\label{sec:common_elements:trav_crane}
The surface building is equipped with a twin-girder 40-tonne capacity overhead travelling crane running on rails attached to the building wall structure. The crane hook has motorised rotation and services supplied to the hook to power the spreader attachment features used to pick up the target, shielding, coil, beam window etc. Cameras and lighting are fitted to the crane and the hook to allow fully remote operation. Recovery in the event of breakdown while handling radioactive loads is ensured by designing in features such as siting the control electronics off the crane and by means of redundancy of drives and cabling.

\subsection{Cooling and ventilation services and equipment}
\label{sec:common_elements:CV}
Cooling and Ventilation (CV) systems for the target complex consist of; water cooling systems for the target, proximity shielding and magnetic coil, a helium purge and purification system for the helium vessel and target, and a pressure cascade ventilation system for the complex to ensure containment of any radioactive contamination. ANSYS\textsuperscript{\tiny\textregistered} simulations have been used extensively to validate the cooling of the target and of the proximity shielding; the results of this work have been fed into the design of the CV systems. Future papers will provide more details of this and of the helium systems. 

The equipment for water cooling and helium purification are housed underground in the target complex along with the sump room to collect any water leaks in the helium vessel. The pressure cascade ventilation ducts will be integrated into the target complex building whilst the rest of the ventilation system equipment and cooling towers will be housed in a separate building to be constructed next to the target complex building.

For the crane concept design, all the cooling and helium system equipment is housed in the underground area (Figure~\ref{fig:Crane_2}); for the trolley concept the cooling and helium systems for the target are installed separately on the rear of the trolley (Figure~\ref{fig:Trolley_hot_cell}).

\section{Crane concept design results}
\label{sec:crane}

\subsection{Overview of the complex}
\begin{figure}[htbp]
\centering %
\includegraphics[width=0.9\linewidth]{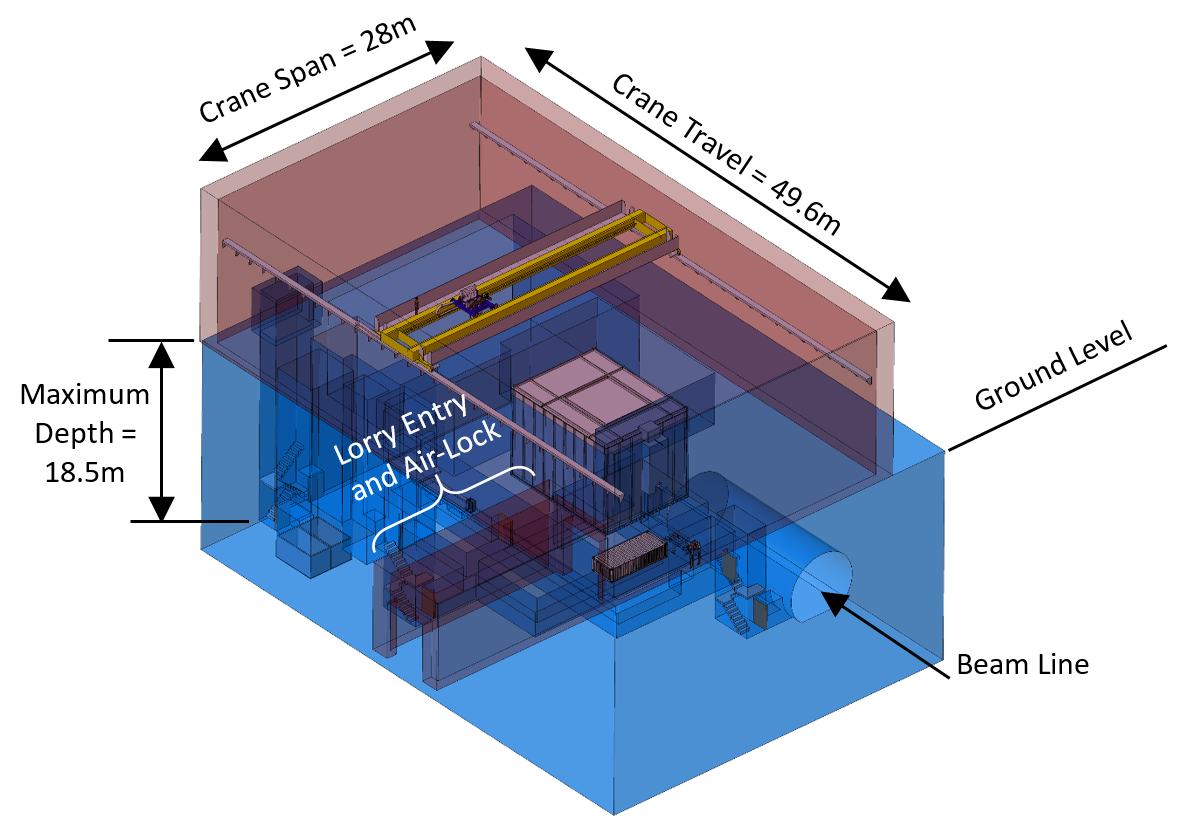}
\caption{\label{fig:Crane_1} Overview of crane concept target complex.}
\end{figure}

\begin{figure}[htbp]
\centering %
\includegraphics[width=0.65\linewidth]{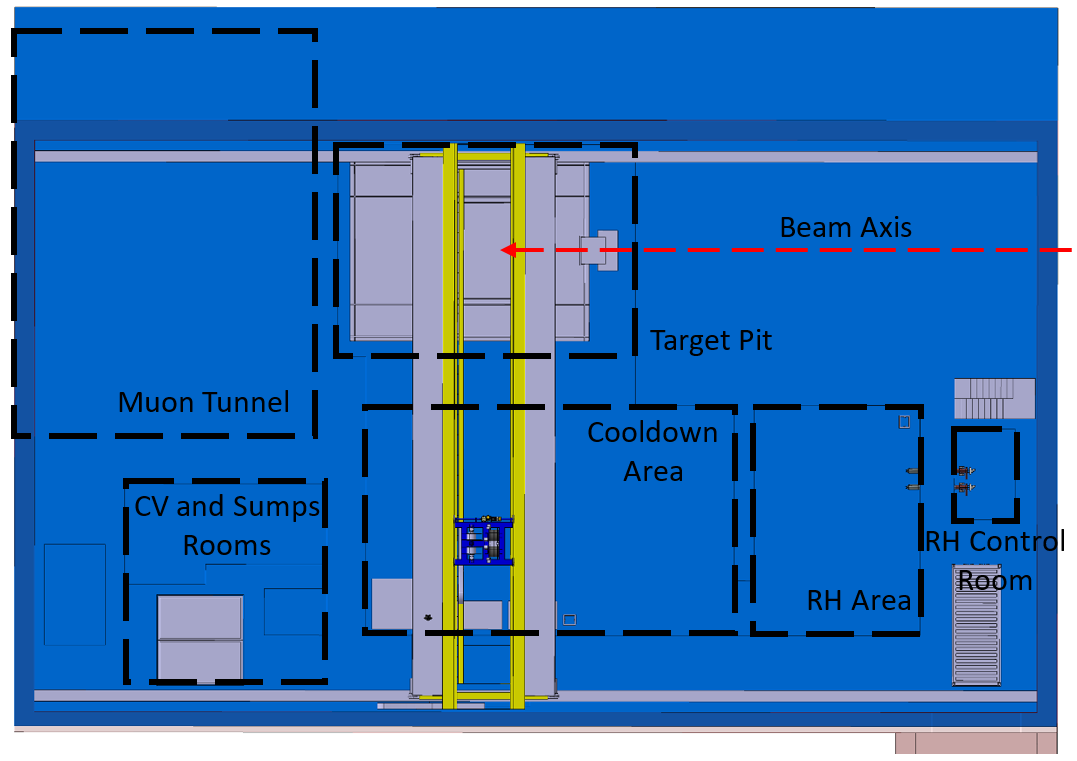}
\caption{\label{fig:Crane_2} Plan view of crane concept target complex.}
\end{figure}

The crane concept target complex design is shown in Figures~\ref{fig:Crane_1} and \ref{fig:Crane_2}. The target, and its surrounding shielding are housed in a helium vessel in the target pit. An underground cool-down area including a remote handling area is used for storage and dismantling of activated components. Cooling and ventilation plant is also underground and sump rooms are used to collect any water leaking from the equipment in the helium vessel. A surface building equipped with an overhead travelling crane covers the underground areas. A vehicle airlock is attached to the surface building.

\subsection{The target, shielding, etc. in the helium vessel}
\begin{figure}[htbp]
\centering %
\includegraphics[width=1\linewidth]{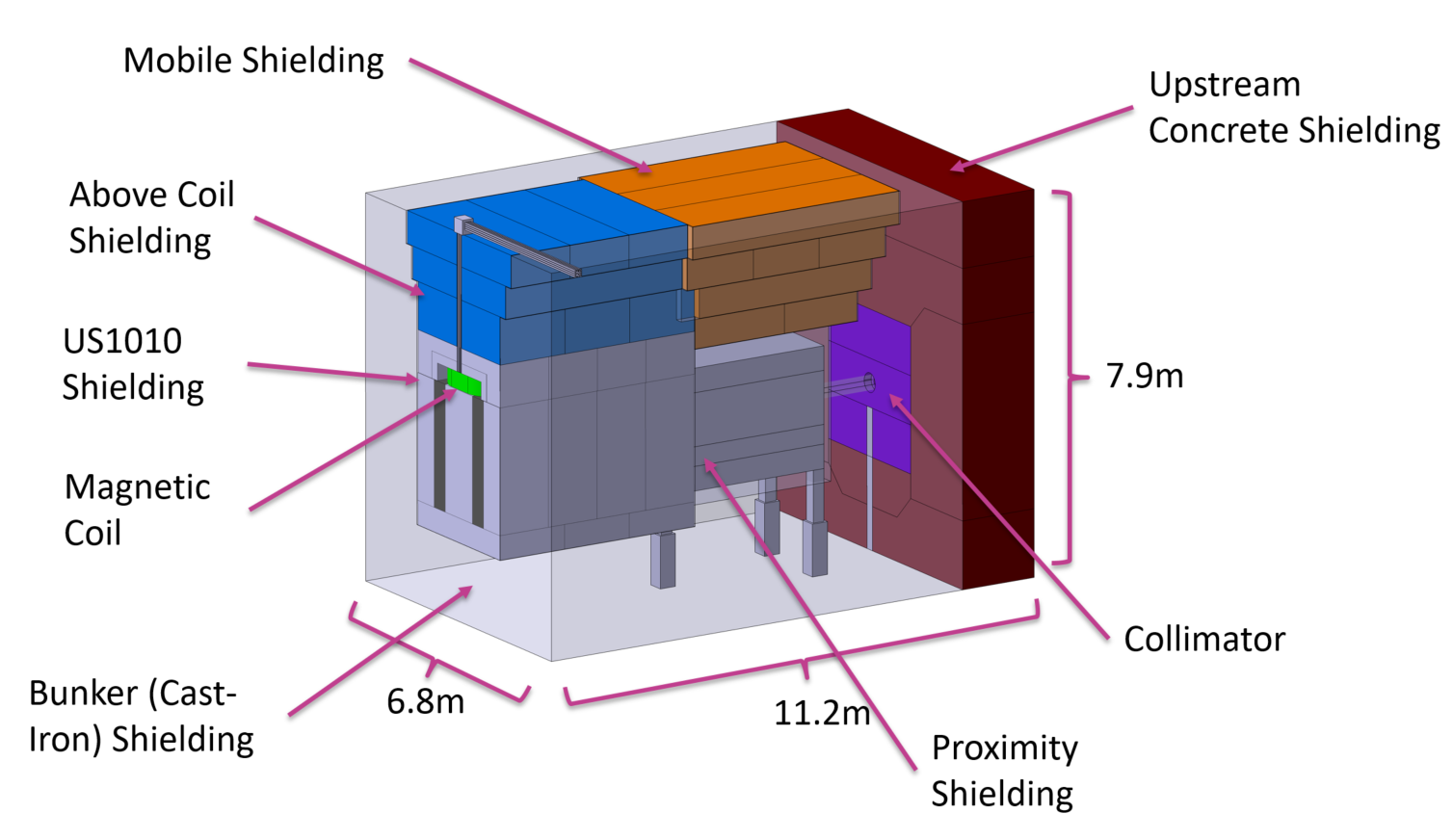}
\caption{\label{fig:Crane_3} Isometric view of crane concept equipment in the helium vessel (bunker shielding and concrete shielding at sides of vessel removed for clarity).  The target is inside the proximity shielding.}
\end{figure}

In the crane concept all installation and removal operations are carried out vertically using the building's overhead travelling crane. In order to exchange the target, it is necessary to first remove the helium vessel small lid (as depicted in Figure~\ref{fig:He_vessel_services}), the ``mobile shielding'' and the top layer of proximity shielding (Figure~\ref{fig:Crane_3}).

\subsection{The target housing, supports, services and connections}
The crane concept target is housed in a rectangular stainless-steel container which incorporates alignment features, lifting points, handling guides, water and electrical connectors (Figure 16). Internal pipework connects the target cooling channels to the remotely operated connector clamps near the base of the target container. The details of the internal pipework are covered by a separate study. 

\begin{figure}[htbp]
\centering %
\includegraphics[width=0.7\linewidth]{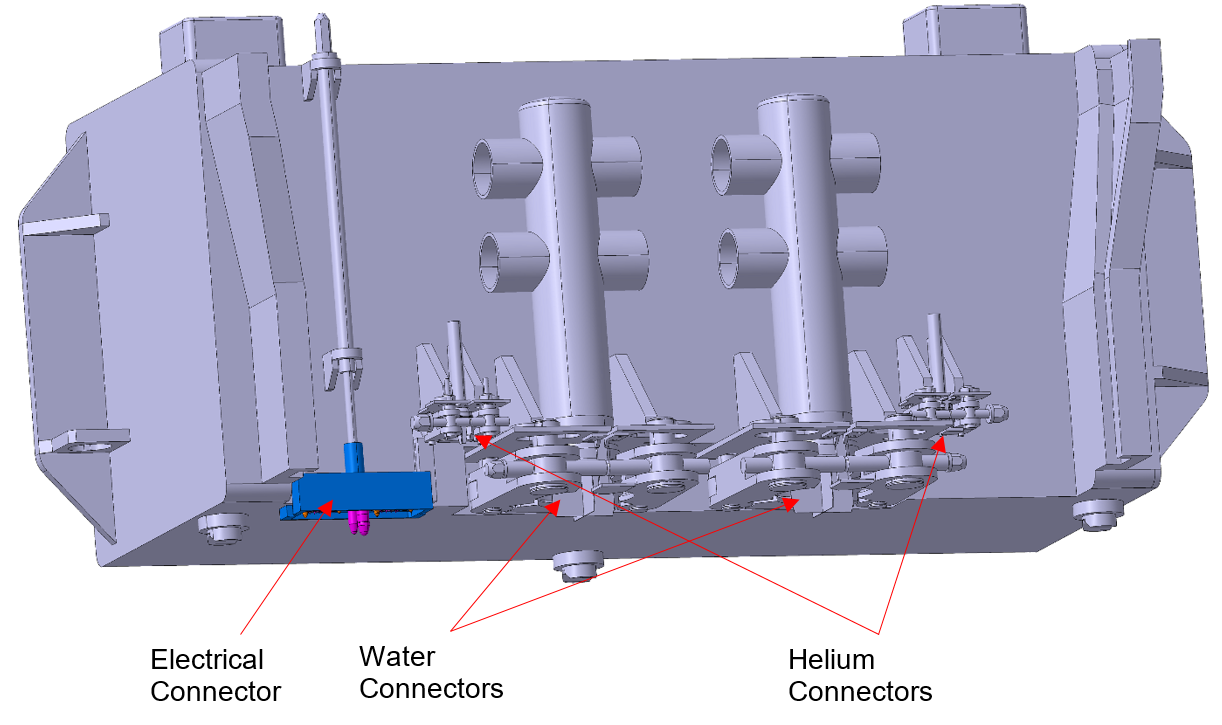}
\caption{\label{fig:Crane_housing_1} Target container showing the service connections (crane concept).}
\end{figure}

The target is supported on the bottom layer of the proximity shielding, which sits on top of three pre-aligned pillars. Cooling water pipework and temperature sensor cables to the proximity shielding pass through the pillars supporting the proximity shielding. A separate service pillar supplies the target with its water, helium and electrical connections.

\begin{figure}[htbp]
\centering %
\includegraphics[width=0.7\linewidth]{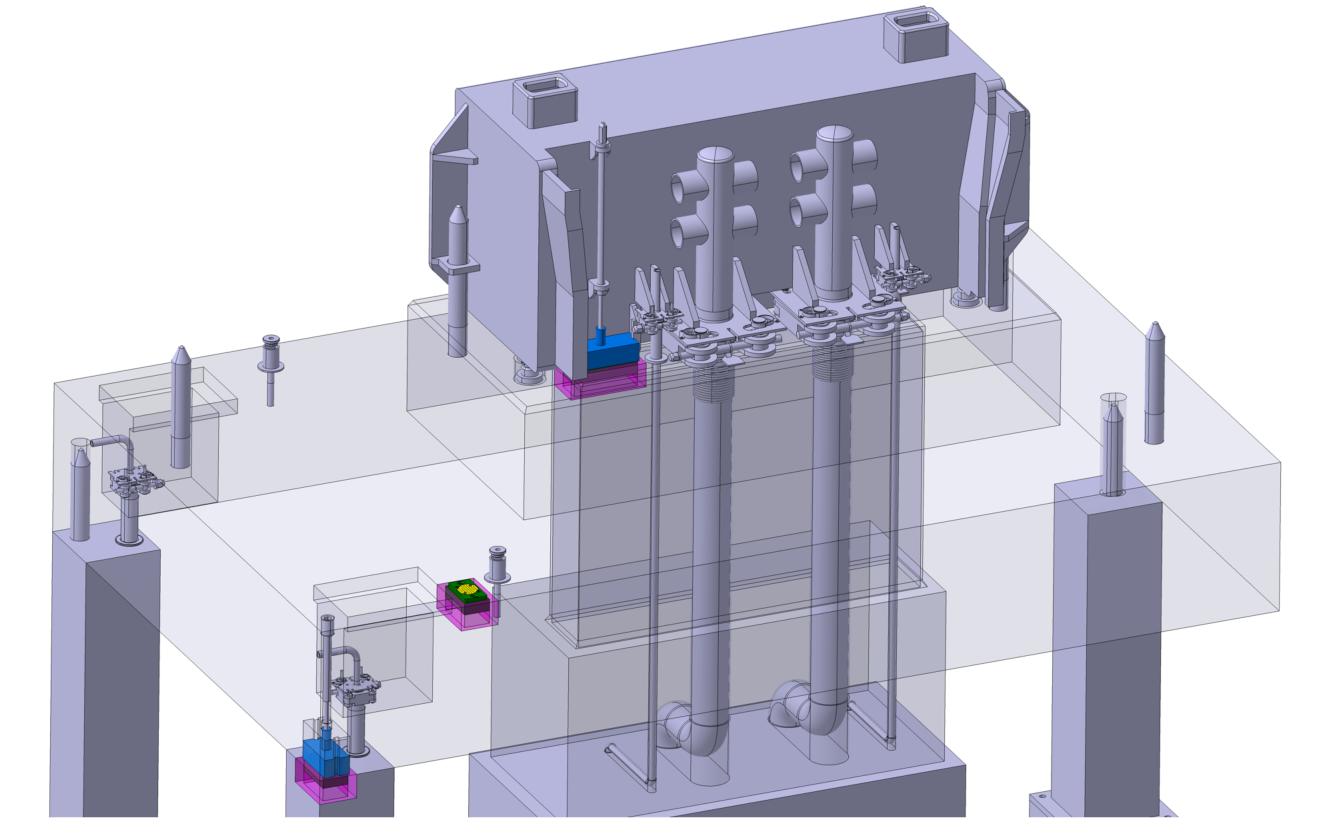}
\caption{\label{fig:Crane_housing_2} Target sitting on bottom layer of proximity shielding (crane concept). Water cooling and electrical connections for the proximity shielding pass through the three support pillars. Pipework connections are made by remotely operated screw-clamp connections.}
\end{figure}

\begin{figure}[htbp]
\centering %
\includegraphics[width=0.65\linewidth]{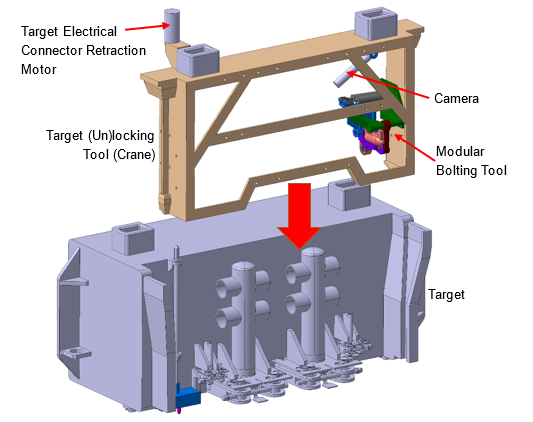}
\caption{\label{fig:Crane_housing_3} (Un)locking tool for target remotely operated screw-clamp water and helium pipework connections (crane concept). Electrical connections are also connected using a motor drive on the (un)locking tool. The tool is lowered into position by the crane. Guidance rails on the target ensure the correct position of the tool.}
\end{figure}

Water connections to the proximity shielding, and water and helium connections to the target, are made using remotely operated clamp-type connections (Grayloc\textsuperscript{\tiny\textregistered} connectors were used for the design) (Figure~\ref{fig:Crane_housing_2}). Electrical connections are made remotely by radiation-tolerant, custom designed connectors once the proximity shielding and target are lowered onto their supports. Clamping and unclamping for the water and helium connections to the target are made by an ``(un)locking tool'' lowered into position by the crane (Figure~\ref{fig:Crane_housing_3}).

\subsection{Proximity shielding}
\begin{figure}[htbp]
\centering 
\includegraphics[width=.35\textwidth]{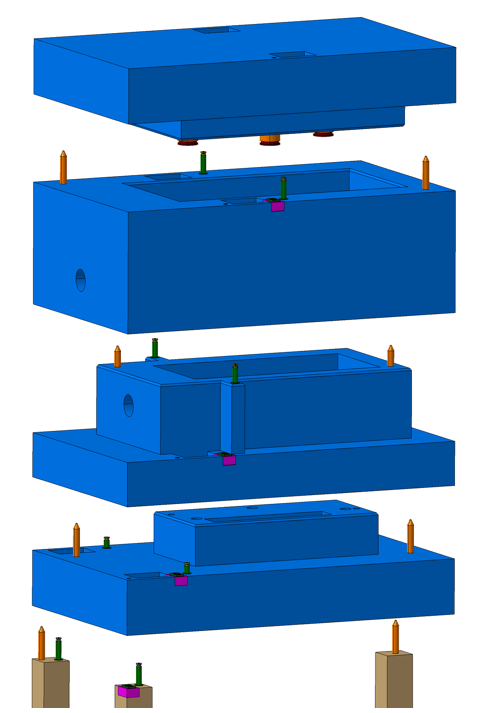}
\qquad
\includegraphics[width=.5\textwidth,origin=c]{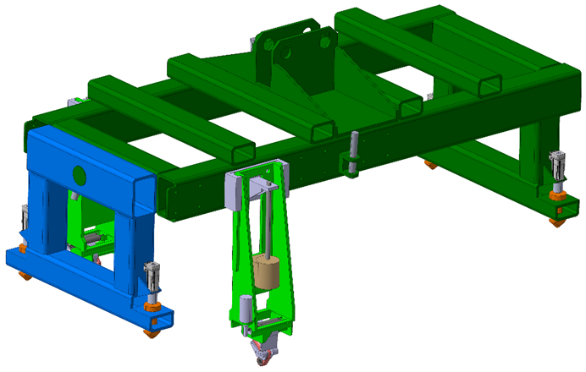}
\caption{\label{fig:proximity_shielding} {\it (Left)} Layers of proximity shielding (crane concept) showing alignment pins and pipework stubs. {\it (Right)} Proximity shielding spreader beam equipped with (un)locking devices for the proximity shielding water connection screw clamps.}
\end{figure}

The crane concept proximity shielding is built up of layers of cast iron with internal cast-in stainless cooling water passages. The coolant pipes are connected using remotely operated screw clamps as used for the target; the connections are made once each layer has been installed on top of the previous layer.  An (un)locking tool, similar to that used for the target, is used to operate the clamps (Figure~\ref{fig:proximity_shielding}).

\subsection{Target exchange}
Removal of the target in the crane concept is carried out by the building overhead travelling crane with a series of remotely operated spreader beams to lift the shielding above the target and then the target itself. The steps involved are given in Table~\ref{tab:target_exchange}. The installation of a new target follows the same procedure but in reverse.

\begin{table}[htbp]
\centering
\caption{\label{tab:target_exchange} The table shows the steps for removal of a target from the helium vessel and transfer to the cool down area (crane concept).}
\smallskip
\begin{tabularx}{\linewidth}{|c|L|L|}
\hline
\textbf{Step} & \textbf{Task} & \textbf{Tooling}\\
\hline
a & Open lid of helium vessel & Hands-on operation \\
b & Remove mobile shielding above target and transfer to cool down area & Crane and remotely operated spreaders\\
c & Disconnect water connections to top layer of proximity shielding & Crane and spreader with (un)locking tool\\
d & Remove top layer of proximity shielding and transfer to cool-down area & Crane and spreader with (un)locking tool\\
e & Disconnect water connections to target & Crane and (Un)locking tool\\
f & Lift out target with shielded spreader and transfer to cool down area & Crane and target transfer spreader (shielded)\\
\hline
\end{tabularx}
\end{table}

\begin{figure}[htbp]
\centering %
\includegraphics[width=0.6\linewidth]{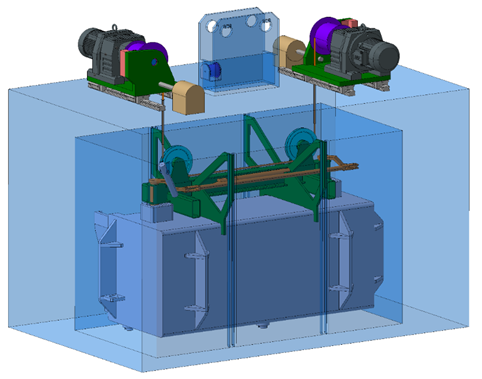}
\caption{\label{fig:target_spreader} Target transfer spreader for shielded transfer of target from helium vessel to cool down area (crane concept).}
\end{figure}

A shielded spreader is used for the target and is equipped with a hoist to lift the target inside the shielding before the target is transferred to the cool down area.  This transfer is carried out with the spreader kept as close as possible to the floor to minimise external radiation shine from the bottom of the target. Additional concrete shielding blocks may be installed along the transfer path to further reduce radiation in the target hall. The target transfer spreader is shown in Figure~\ref{fig:target_spreader}.

\section{Trolley concept design results}
\label{sec:trolley}

\subsection{Overview of the complex}
\begin{figure}[htbp]
\centering %
\includegraphics[width=0.85\linewidth]{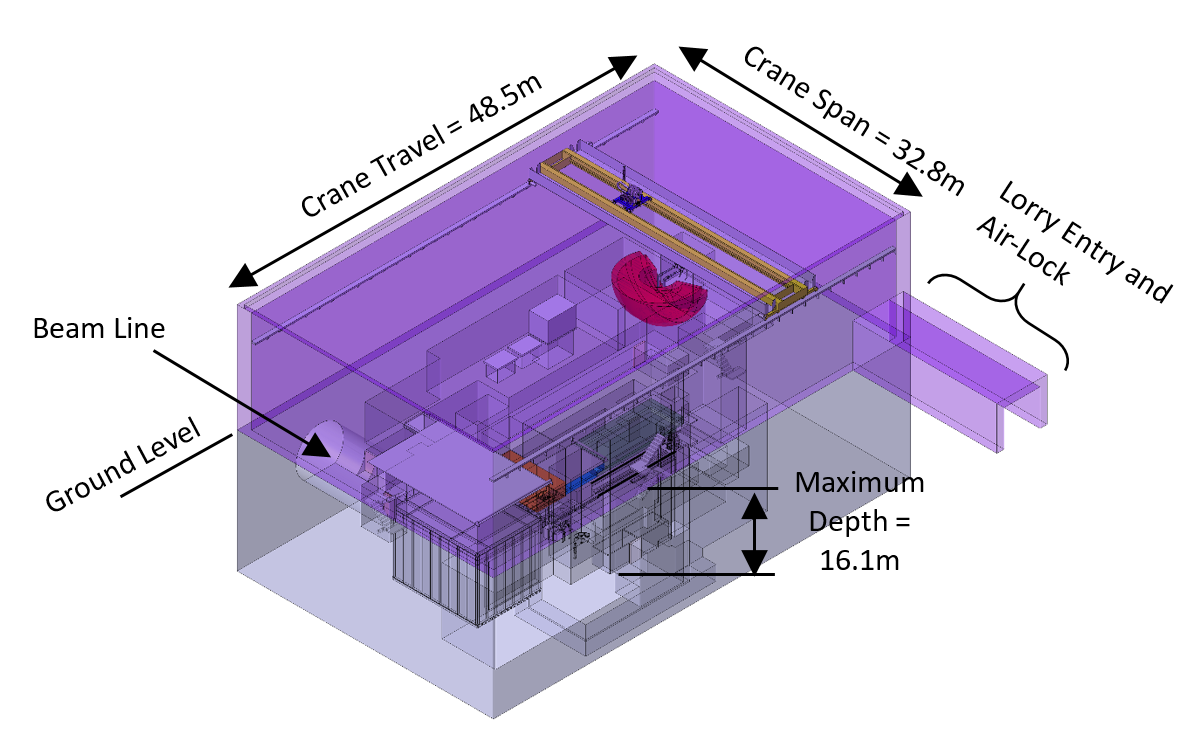}
\caption{\label{fig:Trolley_1} Overview of trolley concept target complex, above and underground areas.}
\end{figure}

\begin{figure}[htbp]
\centering %
\includegraphics[width=0.7\linewidth]{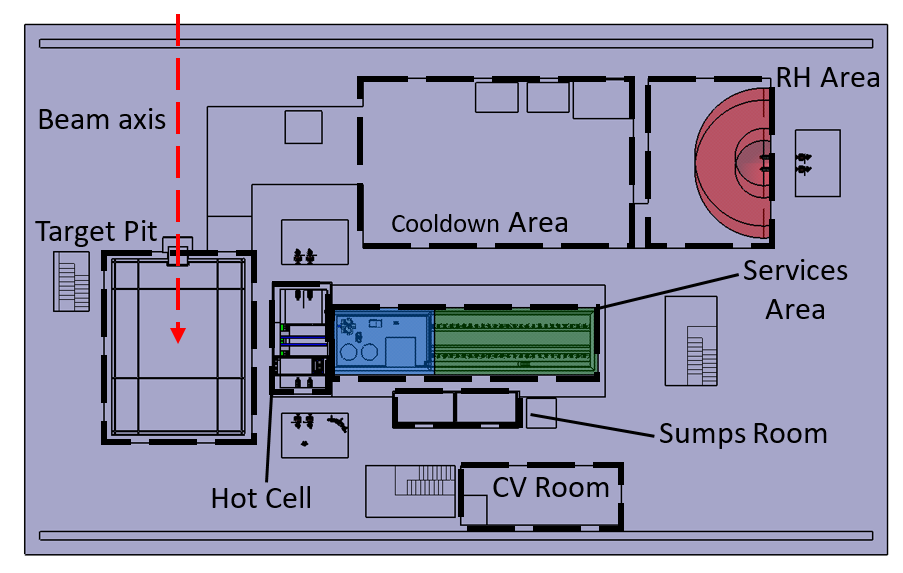}
\caption{\label{fig:Trolley_2} Plan view of trolley concept target complex.}
\end{figure}

The trolley concept target complex design is shown in Figure~\ref{fig:Trolley_1} and \ref{fig:Trolley_2}. As for the crane concept (Section~\ref{sec:crane}), the target and its surrounding shielding are housed within a helium vessel in the target pit. The main difference is that the target is supported on a trolley running on rails that enters the helium vessel from the side (Figure~\ref{fig:Trolley_iso}); this allows the target to be withdrawn from the helium vessel into a hot cell for exchange or repair. The complex includes also complementary underground areas for services and support.

\subsection{The trolley, target, shielding , etc. in the helium vessel}
\begin{figure}[htbp]
\centering %
\includegraphics[width=0.85\linewidth]{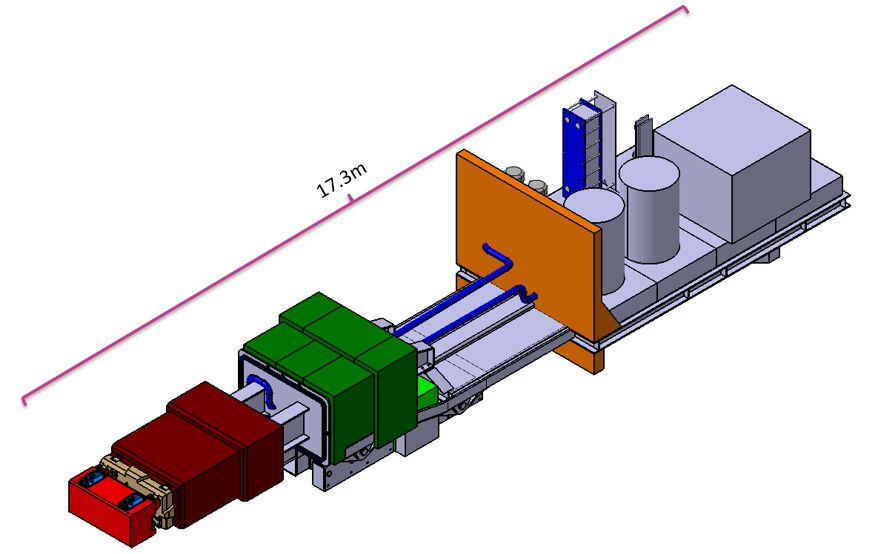}
\caption{\label{fig:Trolley_iso} Overview of trolley assembly. The target is the red assembly on the lower left side, with a several meter thick shielding (cast iron (red) plus concrete (green)) separating it from the hot-cell equipment.}
\end{figure}

The trolley supports the target on cantilever beams which also support a ``plug'' of shielding that moves into the helium vessel along with the target. All the services to the target, and the trolley-mounted water-cooled proximity shielding, are incorporated into the trolley. A door equipped with inflatable EPDM seals is fitted to the trolley to close the helium vessel when the target is in its operating position (Figure~\ref{fig:Trolley_iso}).  

\begin{figure}[htbp]
\centering %
\includegraphics[width=0.85\linewidth]{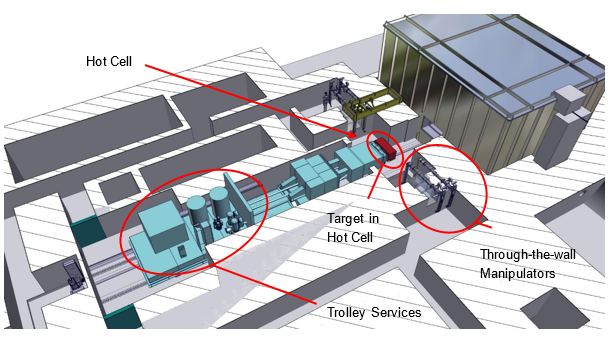}
\caption{\label{fig:Trolley_hot_cell} Cut-away showing trolley withdrawn from helium vessel with target in hot cell.}
\end{figure}

The target can be withdrawn from the helium vessel, without the need to remove the lid of the helium vessel or remove the shielding above the target, by rolling back the trolley. When the target is withdrawn from the helium vessel it enters a hot cell equipped with a 3 tonne overhead travelling crane and two pairs of master-slave manipulators which are primarily used to make and break water and electrical connections to the trolley in the event of target exchange (Figure~\ref{fig:Trolley_hot_cell}).

\subsection{The target housing, supports, services and connections}
\begin{figure}[htbp]
\centering %
\includegraphics[width=0.8\linewidth]{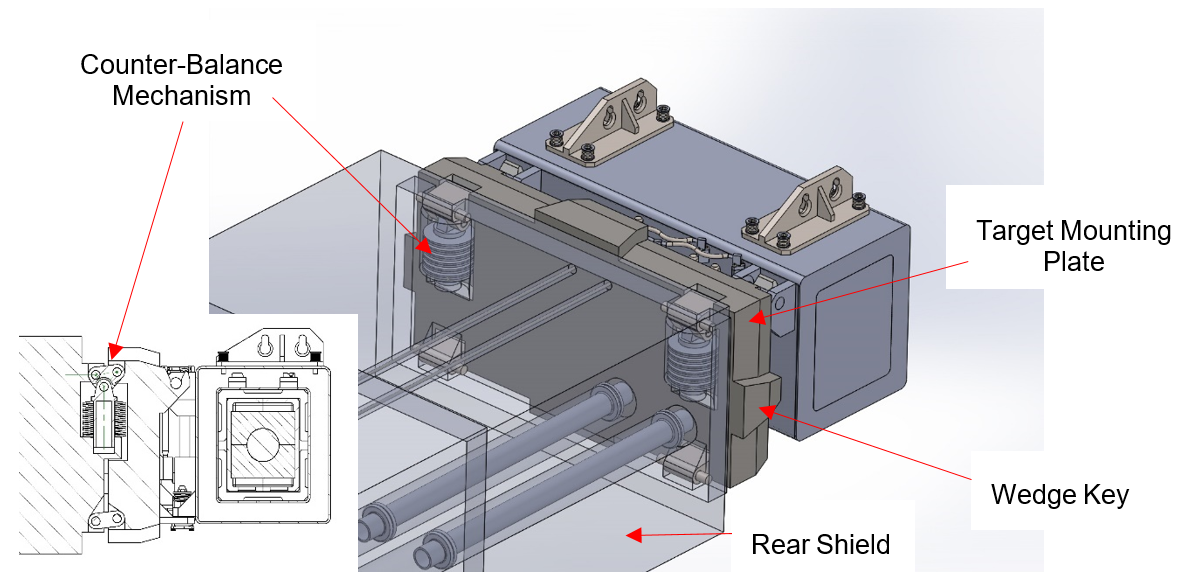}
\caption{\label{fig:Trolley_target_support} Target supported on nose of trolley. Coolant pipes pass through the shielding to the connectors on the side of the target housing.}
\end{figure}

The trolley concept target is housed in a rectangular stainless-steel container which incorporates the support interfaces with the trolley, lift attachment points, water, helium and electrical connectors (Figure~\ref{fig:Trolley_target_support}). Internal pipework connects the target cooling channels to the remotely operated connector clamps on the side of the target container. The details of the internal pipework are covered by a separate study. 

\begin{figure}[htbp]
\centering 
\includegraphics[width=.5\textwidth]{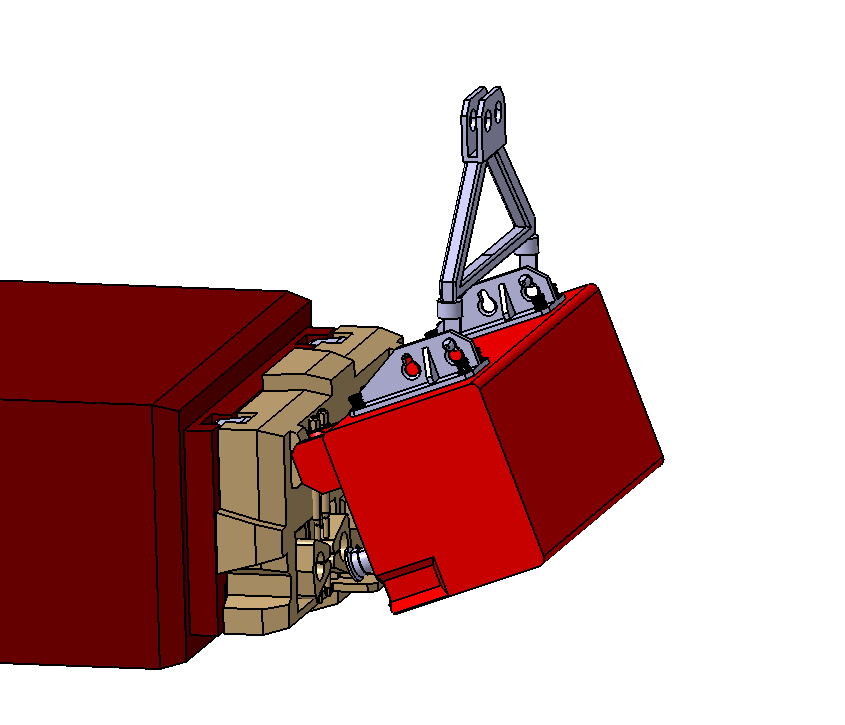}
\qquad
\includegraphics[width=.5\textwidth]{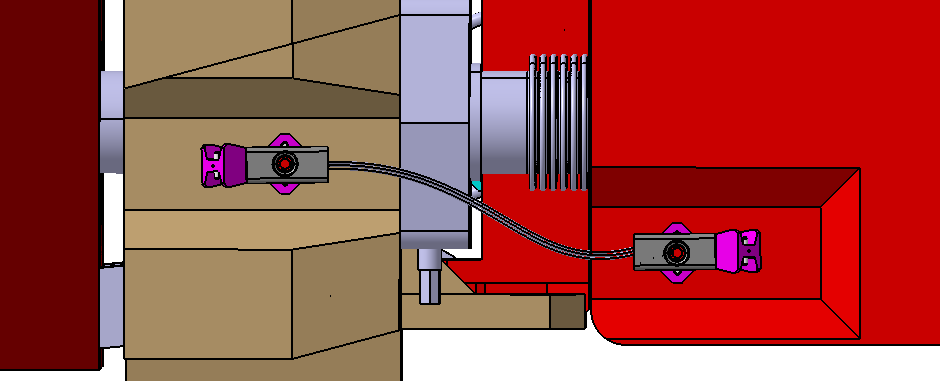}
\caption{\label{fig:Trolley_target_support2} {\it (Top)} BDF target being lowered onto support hooks at the end of the trolley.  Once the target is engaged on the hooks, further lowering rotates the target so that the water connections on the side of the target engage with the screw operated connection flanges on the nose of the trolley. {\it (Bottom)} Electrical connections handled by the hot cell manipulators}
\end{figure}

The target is supported on hooks at the ``nose'' end of the trolley. It is lowered onto the hooks by the hot cell crane and guided into position by the master-slave manipulators (Figure~\ref{fig:Trolley_hot_cell}). All the service connections to the target arrive at the nose of the trolley. Once the target is fully lowered and the water connection pipes are engaged, the screw clamp water connections are tightened by the hot-cell manipulators. Electrical and helium connections are made using the hot-cell manipulators (Figure~\ref{fig:Trolley_target_support2} for the electrical connections).

\subsection{Proximity shielding}
\begin{figure}[htbp]
\centering %
\includegraphics[width=0.9\linewidth]{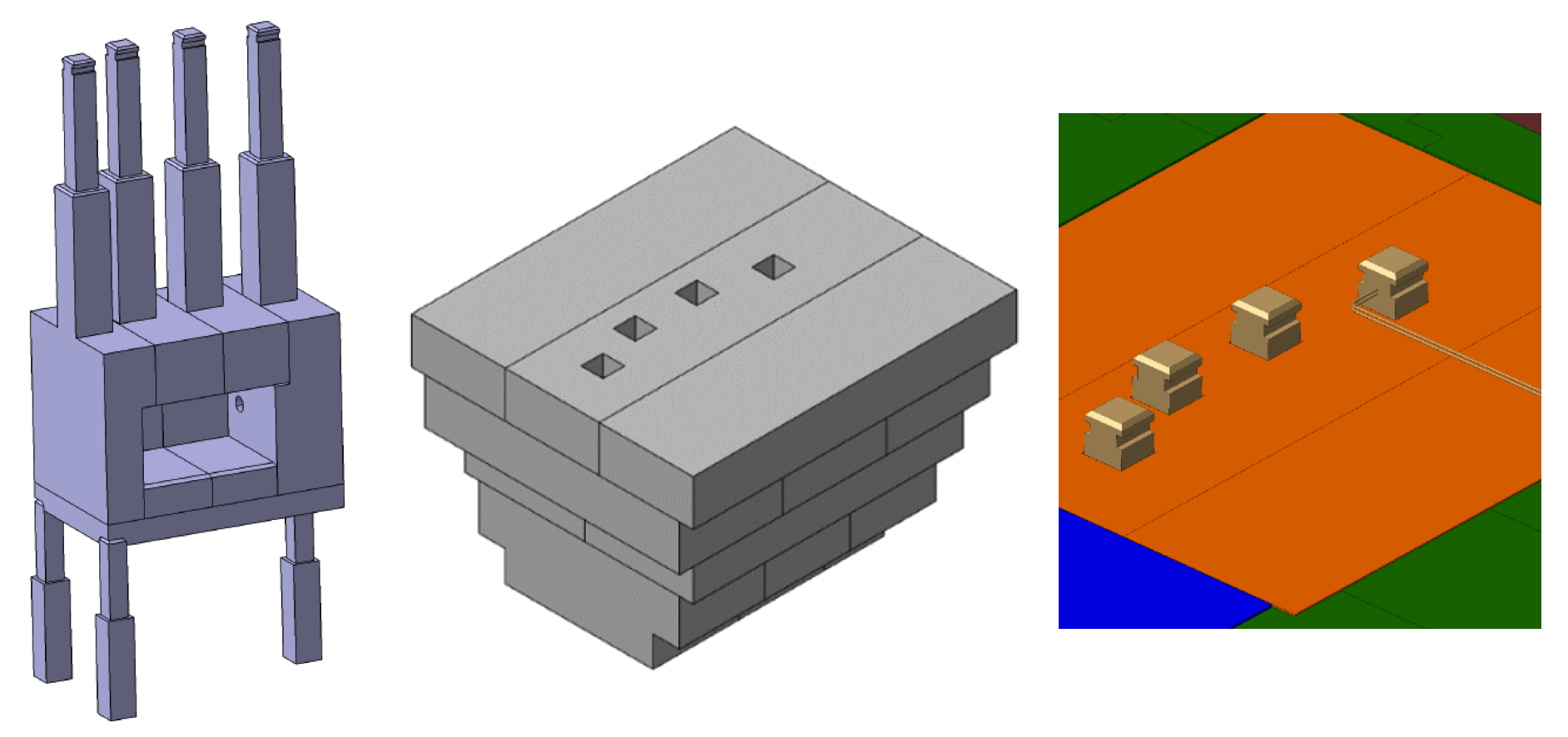}
\caption{\label{fig:Trolley_chimneys} Proximity shielding for trolley concept. {\it (Left)} Four proximity shielding blocks with their service ``chimneys'' (8200~mm above vessel floor).  {\it (Middle)} Mobile shielding above the proximity shielding with passages for the service chimneys (overall dimensions 4775 x 4155 x 3150~mm high). {\it (Right)} Tops of the proximity shielding service chimneys protruding above the mobile shielding with service connections shown on right-hand chimney.}
\end{figure}

The cast iron proximity shielding for the trolley concept is designed to allow the target to enter from the side. It is made up of four elements which include service ``chimneys'' which are used to provide the water cooling and temperature sensor services, while guaranteeing radiation protection shielding. The connections to the proximity shielding are all made at the top of the helium vessel shielding where residual radiation levels are sufficiently low to allow hands-on work. For alignment reasons the proximity shielding is supported on a flat plate which is mounted on support pillars (Figure~\ref{fig:Trolley_chimneys}).

\subsection{Target exchange}
\begin{figure}[htbp]
\centering %
\includegraphics[width=0.9\linewidth]{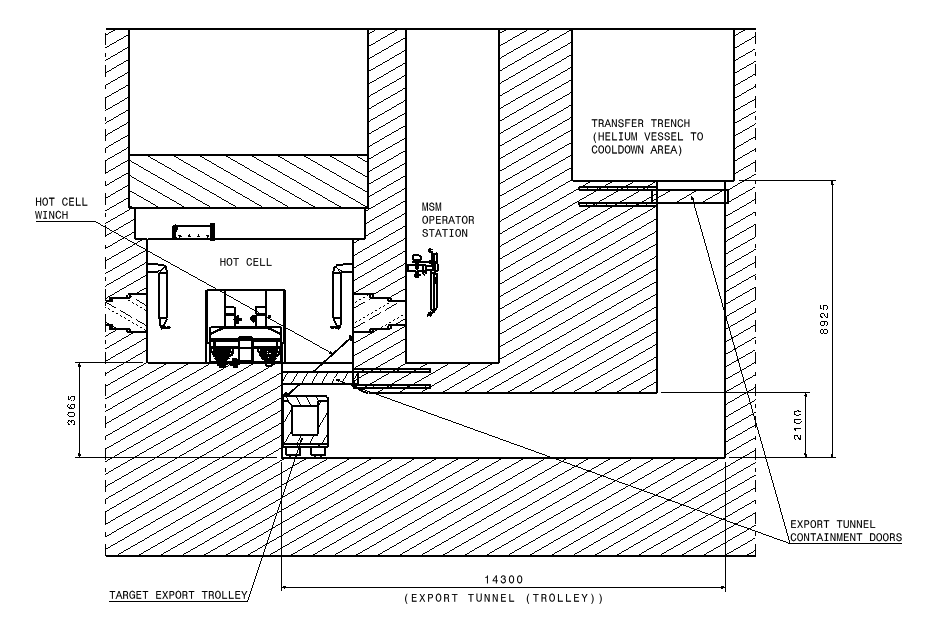}
\caption{\label{fig:Trolley_target_exch} Section through the trolley concept underground area showing the hot cell and target export trolley and the export tunnel linking the hot cell to the cool-down area.}
\end{figure}

Target exchange is carried out by driving the trolley backwards to withdraw it from the helium vessel and into the hot cell.  After disconnection of services in the hot cell using the master-slave manipulators, the target is lifted off the nose of the trolley and placed in a shielded target export trolley using the hot cell crane. The target is then transferred to the cool-down area in the target export trolley that runs along the export tunnel under the hot cell (Figure~\ref{fig:Trolley_target_exch}). Installation of a new target is a reversal of the removal procedure. More details of the steps are given in Table~\ref{tab:trolley_target_exchange}.

\begin{table}[htbp]
\centering
\caption{\label{tab:trolley_target_exchange} The table shows the main steps for removal of a target from the helium vessel and transfer to the cool down area (trolley concept).}
\smallskip
\begin{tabularx}{\linewidth}{|c|L|L|}
\hline
\textbf{Step} & \textbf{Task} & \textbf{Tooling}\\
\hline
a & Withdraw trolley from helium vessel into hot cell & Trolley \\
b & Disconnect and remove target from trolley nose & Hot-cell manipulators and hot-cell crane \\
c & Lower target into export trolley  & Hot-cell manipulators and hot-cell crane \\
d & Transfer target to cool down area & Export trolley, building crane \\
\hline
\end{tabularx}
\end{table}

\section{Recovery from failures - remote handling capability}
\label{sec:recovery}

\begin{figure}[htbp]
\centering %
\includegraphics[width=0.92\linewidth]{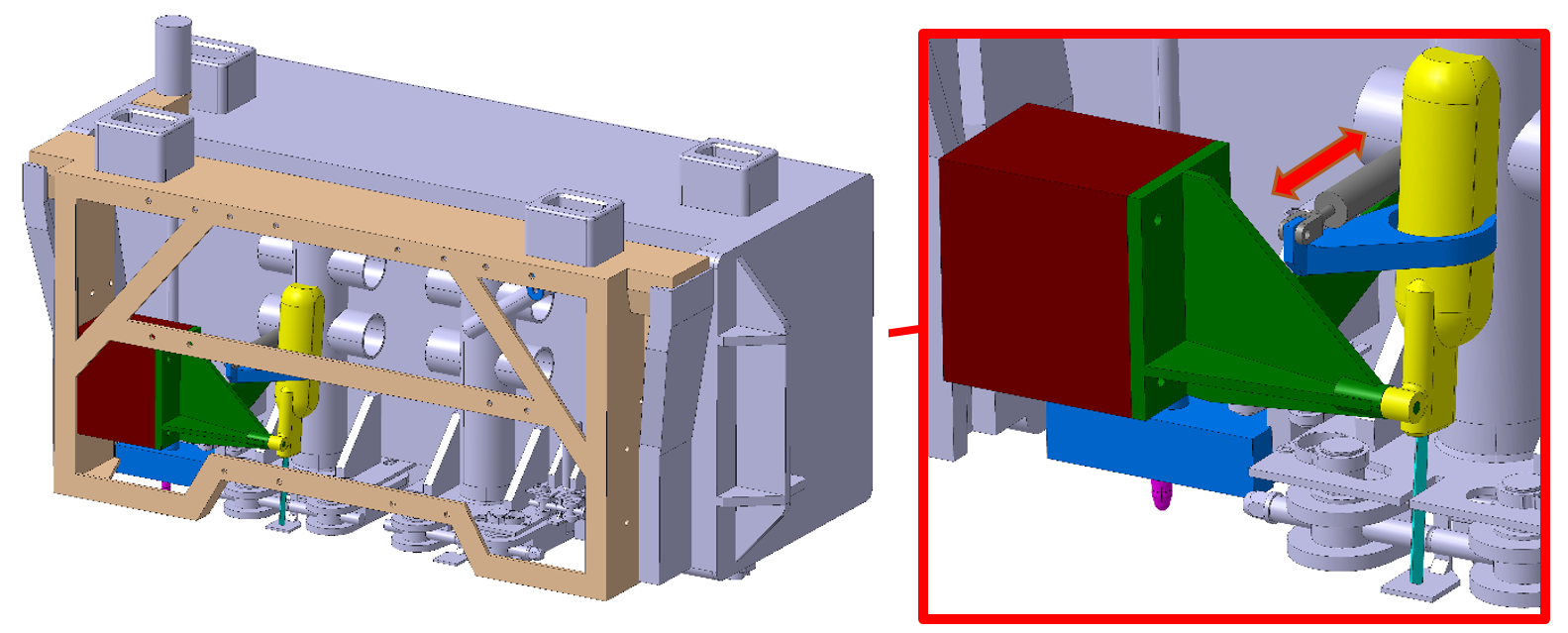}
\caption{\label{fig:unexpected_1} Cutting a seized clamping screw on the target water connection using a saw module installed in the (un)locking tool frame. Saw module shown in zoom view on right.}
\end{figure}

\begin{figure}[htbp]
\centering %
\includegraphics[width=0.75\linewidth]{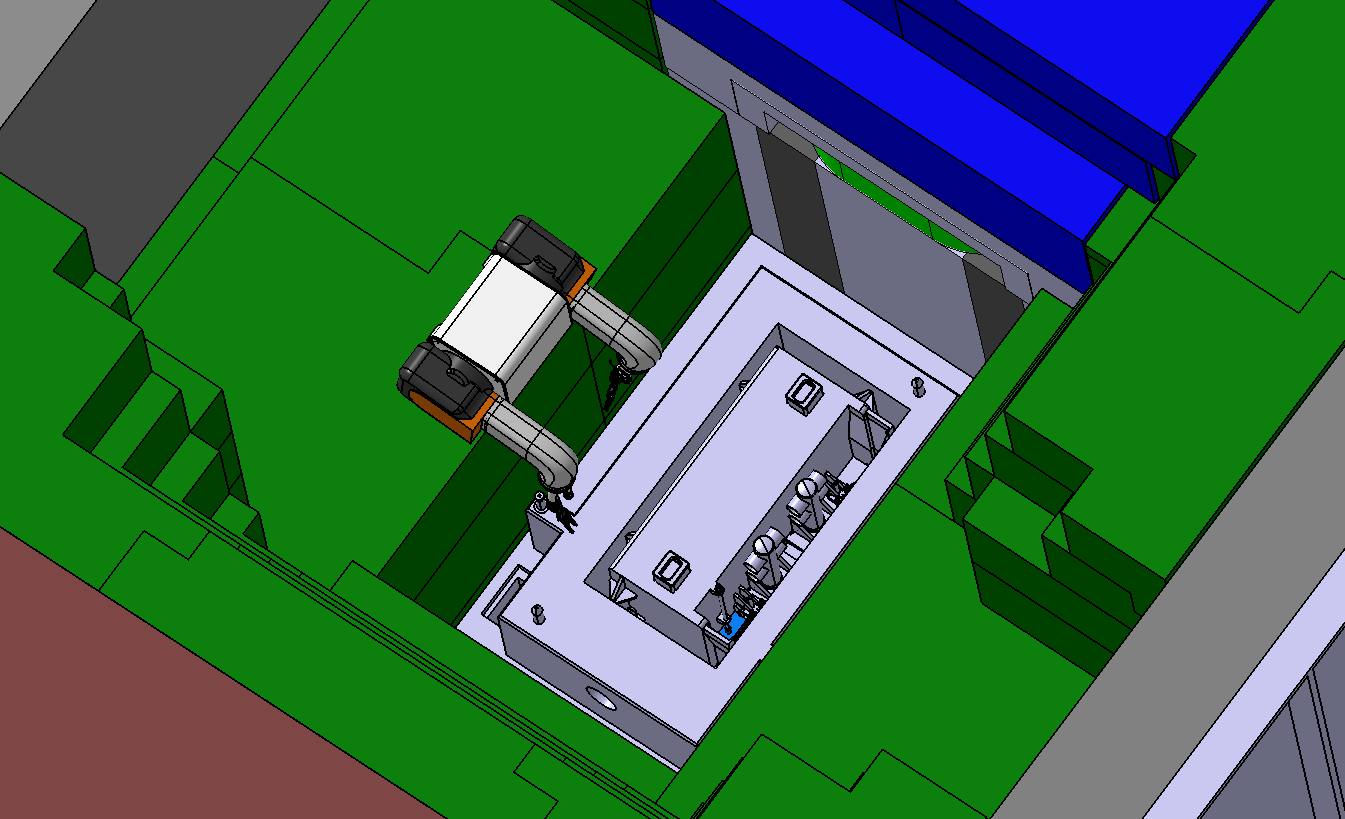}
\caption{\label{fig:unexpected_2} A mobile remote manipulator system for connector repair work in the crane concept helium vessel, showing the target in position with most of the top shielding removed.}
\end{figure}

Recovery from failures was an important aspect of the study due to the very high residual dose rates expected for the target and the proximity shielding. For the trolley concept, failures of connections to the target would be dealt with using the remote handling manipulators and basic ``hand'' tools within the hot cell. For the crane concept there is a need for special tooling to deal with connector failures; a solution based on the use of cutting tools incorporated in the (un)locking tool frame is proposed for this (Figure~\ref{fig:unexpected_1}). However, it was realised that, if the fixed connections in the crane concept helium vessel are damaged, a mobile remotely operated manipulator system would be needed for repair (Figure~\ref{fig:unexpected_2}).

\section{Conclusions and further work}
\label{sec:conclusions}

Both concepts can be used as a basis for the final BDF target complex design. Analyzing the handling and remote handling operations needed during the lifetime of the facility (including those to recover from failures and damage) has led to a clearer understanding of the design requirements for the target complex; the designs produced form a sound basis for further work as the BDF design study advances.

The trolley concept design offers the key advantages of a simpler and faster target exchange along with the flexibility provided by the hot cell which will allow the repair of failed or damaged connectors. The trolley concept has a second advantage of not having fixed water connections in the helium vessel shielding which are a potential source of failures and, in the crane concept design, require the removal of many components to access them. The disadvantage of the trolley concept is the additional development work needed for the trolley itself and the additional space required for it to operate. 

The crane concept offers the advantage of a simpler facility design than the trolley concept. Nevertheless, the tooling required to operate and recover from failure of the water connections in the helium vessel will require extensive development and testing. For the crane concept, repair of damaged water or electrical connections in the helium vessel will require an additional mobile remote handling manipulator system.

Further design work is underway to develop the BDF design, such as the target (including supports and connections), the helium vessel, the civil engineering and services such as cooling and ventilation, helium purification, electrical supplies and integration of the complete facility on the CERN site.



\acknowledgments
The authors would like to thank personnel from the various CERN services that contributed to the design and the design reviews held during the study and to thank J.P. Dauge for providing the maps used. In addition, the authors would like to thank G. Gilley and J. Boehm from the Rutherford Appleton Laboratory for their inputs on the magnetic coil and US1010 shielding design.
The work benefitted from (support and) close cooperation with the (whole) SHiP Collaboration.
The authors kindly thank the Rutherford Appleton Laboratory ISIS Target team for the support given in the development of the ``trolley'' concept. MC kindly acknowledges the technical exchanges with the T2K Target Station team at J-PARC and with the LBNF Project Secondary beamline design team. 


\bibliographystyle{unsrtnat}
\bibliography{BDFreferences}

\end{document}